**Title:**

# MicroCup: A Cryogenic Specimen Preparation Strategy for Atom Probe Tomography of Organic Molecular Liquids

**A short running head (<50 characters):**

MicroCup enabled APT of organic molecular liquids

**Authors:**

Kuan Meng[1], Florian Groll[1], Sebastian Eich[1], Guido Schmitz[1]

**Affiliations and address of all authors:**

1) University of Stuttgart, Institute for Materials Science, Heisenberg Str. 3, 70569 Stuttgart, Germany

**Primary institutions where the research was performed:**

University of Stuttgart, Institute for Materials Science, Heisenberg Str. 3, 70569 Stuttgart, Germany

**Corresponding Author Email Address:**

Kuan.Meng@imw.uni-stuttgart.de

**Mailing address:**

University of Stuttgart, Institute for Materials Science, Heisenberg Str. 3, 70569 Stuttgart, Germany

**Telephone number:**

+49 711 685-61961

**Fax number:**

+49 711 685-51902

**Conflict of interest statements for all authors：**

The authors declare that they have no known competing financial interests or personal relationships that could have appeared to influence the work reported in this paper.

**Abstract:**

Atom probe tomography (APT) of organic molecular liquids is limited by poorly reproducible specimen geometry, reduced milling rates, and beam sensitivity during cryo-FIB preparation. Here we introduce a MicroCup strategy that confines liquids in a FIB-prepared nanoscale cavity prior to phase separation, reduces deposited volume to increase preparation throughput, enables reproducible specimen geometry, and minimizes beam exposure in the region of interest. Using the liquid crystals 4′-octyl-4-cyanobiphenyl (8CB) and 4′-octyloxy-4-cyanobiphenyl (8OCB) as model systems, we establish stable and reproducible field evaporation conditions, enabling the detected intact ion molecular preservation above 70% in smectic-like phases with interpretable fragmentation behavior. Comparative analysis further shows that the oxygen atom in 8OCB promotes preferential cleavage pathways associated with bond polarization under high electric fields. By inducing partial crystallization within the MicroCup cavity, distinct regions could be resolved: 8CB shows broadly similar evaporation behavior across crystalline and amorphous regions, whereas 8OCB exhibits clearer regional contrast, with smectic-like regions dominated by intact molecular or large fragments and crystalline domains producing small alkyl fragments and ether-type species. These results provide spatially resolved evidence of a solid-liquid interface in a freeze-prepared organic liquid by APT and establish a reproducible workflow for probing local phase behavior in soft materials.

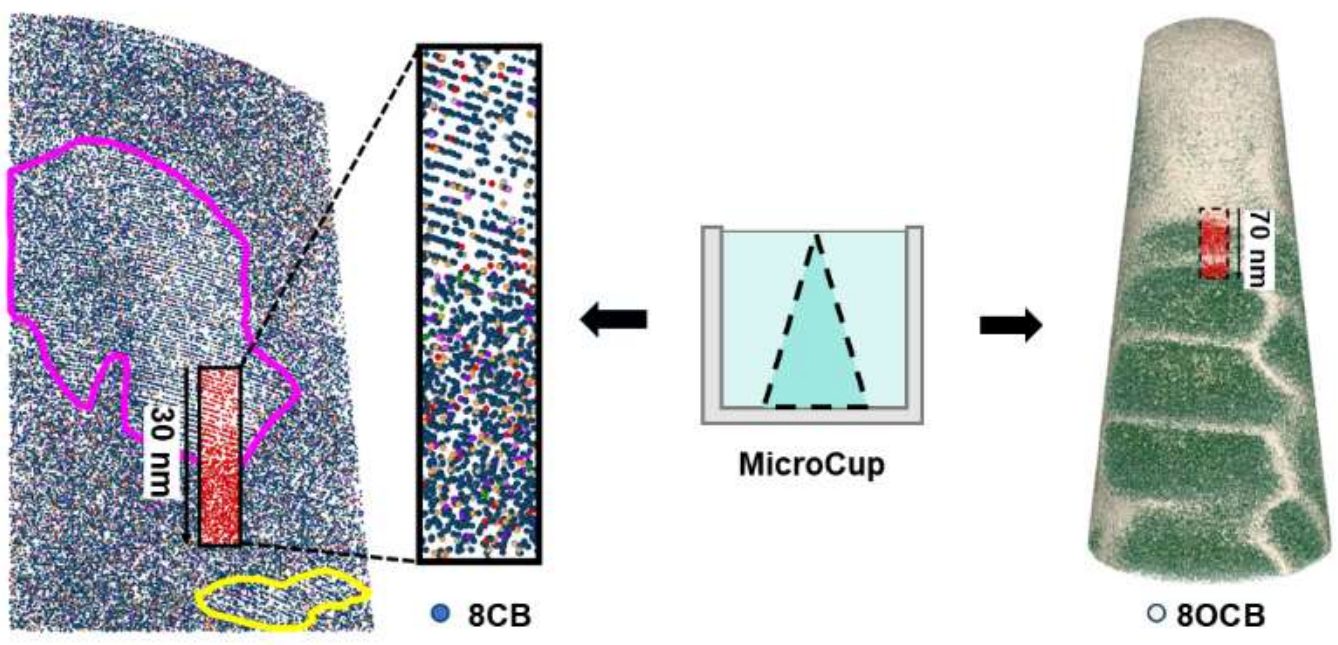

**Keywords:** atom probe tomography, cryo-focused ion beam, soft matters, molecular liquids, liquid crystals, cyanobiphenyls.

## 1. Introduction

Organic molecular liquids represent an important yet largely unexplored frontier for atom probe tomography (APT) (Devaraj et al., 2018). Despite their central role in soft matter systems, from liquid crystals (LCs) (Bukusoglu et al., 2016) and electrolytes (Li et al., 2023) to catalytic systems(Dyson & Jessop, 2016), their analysis by APT remains limited by a fundamental challenge in specimen preparation. APT requires a mechanically stable specimen with a needle-shaped geometry, whereas liquids do not possess an intrinsic shape. For organic molecular systems, this challenge is further compounded by the need to preserve chemically meaningful molecular information during specimen preparation. Unlocking this capability would expand the practical boundary of APT for organic materials and deepen our understanding of prior APT studies involving organic components (Nishikawa & Kato, 1986; Nishikawa et al., 2009; Gordon & Joester, 2011; Stoffers et al., 2012; Ji et al., 2020; Rivas et al., 2020; Meng et al., 2022; Van Vreeswijk et al., 2022; Pantenburg et al., 2023).

In recent years, the APT community has made significant efforts in analyzing aqueous liquids using cryogenic transfer and cryogenic focused ion beam (FIB) workflows (Schreiber et al., 2018; Qiu et al., 2020; El-Zoka et al., 2020; Schwarz et al., 2020, 2021, 2022; Stender et al., 2022; Zhang et al., 2022; Exertier et al., 2024; Tegg et al., 2024; Woods et al., 2023, 2025; Zhang et al., 2025). These efforts highlight a growing interest in extending APT toward soft and beam-sensitive materials (Narayan et al., 2012; Greiwe et al., 2014; Pfeiffer et al., 2017; Kim et al., 2022; Katnagallu et al., 2025). However, the workflows developed for aqueous systems do not straightforwardly translate to organic liquids. Compared to water-based droplets, organic materials exhibit substantially more pronounced beam-induced deformation (Kim et al., 2011), altered surface chemistry under ion irradiation (Briggs & Hearn, 1986; Lee, 1999; Bassim et al., 2012; Zhang et al., 2025), and significantly reduced milling rates (Mihara et al., 2017; Tiddia et al., 2020), all of which fundamentally limit their compatibility with existing specimen preparation approaches.

Current strategies for preparing organic materials for APT can be broadly grouped into two routes. One relies on deposition-based methods (Gault et al., 2010; Zhang & Hillier, 2010; Prosa et al., 2010; Stoffers et al., 2012; Proudian et al., 2016; Eder et al., 2017; Proudian et al., 2019; Bingham et al., 2021; Solodenko et al., 2022), which avoids direct beam exposure and preserves molecular integrity. However, deposition is intrinsically better suited to thin surface films rather than free-standing bulk liquids. Moreover, fragmentation behavior can depend strongly on the geometry of the final APT specimen (Prosa et al., 2010). This dependence is most likely indirect, as specimen geometry controls the local electric field, while the field itself is the key parameter governing molecular fragmentation. Therefore, reproducible control over final specimen geometry is essential not only for obtaining reproducible mass spectra, but also for systematically understanding molecular field-evaporation behavior.

Alternatively, FIB-based approaches provide excellent control over specimen geometry but introduce substantial challenges when applied to organic materials. In our preliminary experiments, even a single 3 nA ion-beam snapshot caused noticeable collapse and material migration in organic droplets (Meng et al., 2026). Even under cryogenic conditions that mitigate chemical degradation (Rivas et al., 2020), the ion beam can still drive irreversible deformation and chemistry modification in organic droplets during imaging and shaping. Furthermore, organic materials readily form modified surface layers under ion irradiation (Tiddia et al., 2020), which dramatically reduce the milling rate: preparing a single n-alkane specimen out of a frozen droplet required over 10 hours (Meng et al., 2022), compared to only 4 hours for a frozen water sample from a droplet of comparable size (Schwarz et al., 2020; Stender et al., 2022). This time inefficiency, combined with the requirement to avoid beam exposure, significantly increases the failure rate and hinders systematic studies.

For liquids, the central question is not only how the specimen evaporates, but whether freezing preserves a meaningful liquid-derived volume in the first place. Because APT interrogates a frozen solid, crystallization during quench freezing can erase liquid or mesophase signatures through solute rejection and concentration gradients (Zhang et al., 2025; Woods et al., 2025). Therefore, suppressing crystallization and achieving vitrification (Wowk, 2010) within the analyzed volume is a key prerequisite for interpreting APT data as liquid-derived structural information.

These limitations collectively highlight a fundamental bottleneck: there is currently no method that enables controlled specimen morphology, minimal beam damage, practical throughput and preservation of a vitrified analysis volume for APT analysis of organic molecular liquids.

To address these challenges, we propose a new specimen preparation strategy by confining the liquid inside a FIB-milled MicroCup cavity prior to solidification, followed by controlled FIB shaping of the liquid into a needle-shaped specimen. Related confinement-based approach were reported previously. Li et al. deposited catalyst nanoparticles embedded in an ionic liquid into a sub-micrometer cavity on a Si microtip post at an elevated temperature. Unlike the liquid systems studied here, the ionic liquid used in their study has a melting point of 49 °C and therefore remains solid at room temperature, without requiring cryogenic preparation (Li et al., 2019). To improve the reliability of APT measurements, our method further ensures the frozen apex stands well above any neighboring metal base (**Figure S1.1**). In our trials, specimens near the metal base did not yield molecular signals. This might be because the metal base causes field distortions, hindering stable evaporation from the apex. Combined with minimized beam imaging, this approach reduces beam-induced damage while enabling a well-defined specimen geometry and stable field-evaporation conditions. In addition to improving preparation throughput, the reduced deposited volume and lateral heat extraction provided by the MicroCup geometry increase the chance of analyzing vitrified, liquid-derived material rather than crystallized regions.

We demonstrate these advantages of the MicroCup method using the thermotropic LC 4’-octyloxy-4-cyanobiphenyl (8OCB) as the model system and comparing it to previously studied 4’-pentyl-4-cyanobiphenyl (5CB) and 4’-octyl-4-cyanobiphenyl (8CB) prepared via the conventional approach based on a fractured W post (Meng et al., 2026). Here, nCB denotes an n-carbon alkyl chain (n = 5 or 8) attached to a biphenyl core (B) terminated by a cyano group (C). In particular, 8OCB differs from 8CB by a single oxygen atom linking the alkyl chain to the cyanobiphenyl core. The corresponding molecular structures are shown in **Figure S2.1**. Based on these results, we further establish practical geometric criteria for achieving stable and reproducible field evaporation in organic molecular liquids. Furthermore, this well-defined structural difference enabled a direct comparison of their fragmentation pathways under controlled evaporation conditions and isolates signatures associated with the oxygen site. By inducing partial crystallization within the MicroCup, we show that 8CB and 8OCB exhibit distinct, phase-dependent evaporation behavior in the vicinity of interfacial regions. Overall, these advances turn organic molecular liquids into a practically accessible target for APT, enabling reproducible spectra and phase-resolved evaporation signatures that were previously difficult to obtain.

## 2. Materials and Methods

### 2.1 Materials

8CB (purity: 98%, Merck, Darmstadt, Germany) and 8OCB from (purity: 99.5%, SYNTHON Chemicals, Wolfen, Germany) were investigated in this study. At room temperature, 8CB appears as an opaque liquid while 8OCB exists as white crystalline powder. Using both liquid- and solid-phase cyanobiphenyl derivatives highlights the versatility of the MicroCup method. W wire (diameter: 50 μm, Thermo Fisher Scientific Inc., Waltham, MA, USA) was employed as the material support for MicroCup fabrication.

### 2.2 Specimen Preparation

MicroCups were fabricated directly on electropolished W wires. Prior to FIB processing, the W wires were reduced to a diameter of approximately 20 μm (**Figure 1**a) by electropolishing, following our previous procedure (Duran, 2023). The FIB preparation then consisted of two sequential milling steps. First, the ion beam was directed laterally onto the W wire to flatten its end and expose a cross-sectional surface with a diameter between approximately 15 and 20 μm. This was achieved using a dual-beam focused ion beam/scanning electron microscope (FIB/SEM, Thermo Fisher Scios, Waltham, MA, USA; Ga ion source), with the sample mounted on a stage pre-tilted by 45° and rotated by 180° for rectangular pattern milling. In the second step, the exposed cross section was re-oriented to be perpendicular to the ion beam by rotating the stage back to 0° and tilting it by approximately 7°. A circular milling pattern was then applied to produce the cup-shaped cavity. For the present study, the MicroCup diameter ranged from 10 to 15 μm and the depth was approximately 5 μm. The cavity was milled at 30 kV and 5 nA for approximately 6 min. It should be highlighted that the diameter (X) and depth (Z) of the cavity in this method can be adjusted with nanometer precision by tuning the milling parameters (**Figure 1**b). Alternative milling patterns could also be used to generate different MicroCup geometries when required.

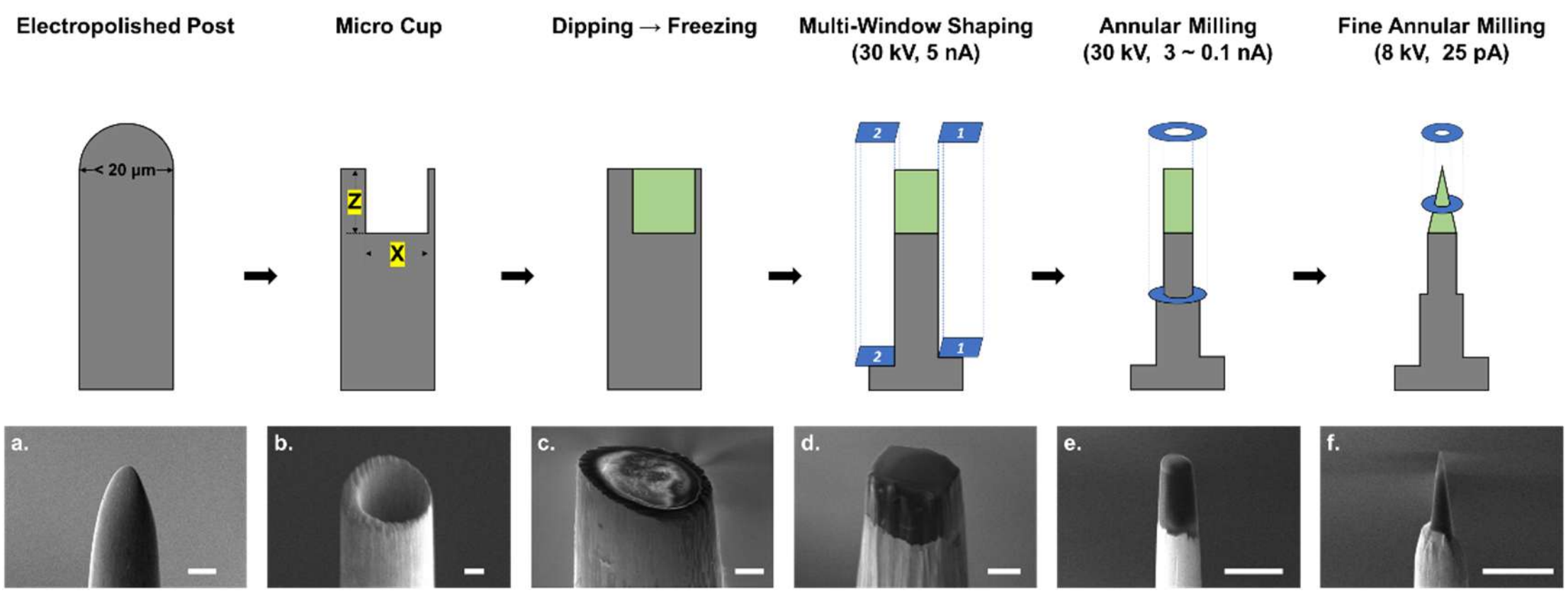

***Figure 1.** Schematic of 8OCB preparation via MicroCup method. X,Z can be controlled via the milling process, where the scale bar is 4 μm.*

For 8OCB deposition, the material was gently heated to 363 K to reach the isotropic liquid state before being filled into the MicroCup using a glass Pasteur pipette (DWK Life Sciences GmbH, Wertheim, Germany). To ensure complete filling without excess accumulation on the surface, the MicroCup and its holder were maintained at the same temperature during deposition. Subsequently, samples were cooled down to room temperature, well below the crystallization transition temperature (330 K) (Dayal et al., 2006), for approximately 30 min prior to quench-freezing by manual plunging into a liquid nitrogen bath using a Leica EM VCM (Leica Mikrosysteme Vertrieb GmbH, Wetzlar, Germany).

After quench-freezing, freeze-etching was performed in a Leica EM ACE600 high-vacuum coater (Leica Mikrosysteme Vertrieb GmbH, Wetzlar, Germany) at 185 K and $10^{-5}$ mbar for 60 min to remove surface water contamination without further evaporating the organic material (**Figure 1**c). Then the sample was mounted on a FIB cryo stage, maintained at 123 K by copper bands connected to a liquid nitrogen dewar. Sample transfer connecting these steps was carried out at 89 K in a Leica VCT500 shuttle (Leica Mikrosysteme Vertrieb GmbH, Wetzlar, Germany) with a vacuum better than 0.02 mbar.

Low-dose beam imaging was used to monitor the milling process and minimize beam-induced damage. Electron imaging was performed with 5 kV and 50 pA (Schwarz et al., 2020). Ion beam imaging of the region of interest was restricted to currents below 0.3 nA to minimize beam-induced damage, whereas higher ion-beam currents were used only for milling the surrounding material. The milling procedure followed the approach described in our previous work. Initially, a multi-window shaping technique (Meng et al., 2022, 2026) was applied to process the cup wall regions. In this step, multiple rectangular patterns are sequentially milled around the specimen's circumference to gradually reduce its size. Once the lateral sample size reached below 10 μm (**Figure 1**d), conventional annular milling (Larson et al., 1998) was employed to smooth edges and further diminish sample size while lowering ion current progressively. Throughout the milling process, a depth of at least 10 um was removed from both the LC and the W substrate to avoid significant secondary protrusions near the frozen organic apex (**Figure S1.1**c-d). When the sample diameter was reduced below 3 μm (**Figure 1**e), final milling was performed using a lower acceleration voltage (Perea et al., 2016; Ji et al., 2020) of 8 kV and 25 pA to minimize Ga implantation. The process was terminated when the inner diameter reached approximately 200 nm (**Figure 1**f).

**2.3 Measurement Conditions**

The prepared specimens were measured in a custom-made atom probe (Schlesiger et al., 2010), equipped with a 120 mm delay line detector system and an open area ratio (OAR) of 50 %. Laser pulse mode was used at a wavelength of 355 nm, with a 250 fs pulse length and a 30-micron spot size. All samples were measured at 60 K, with the laser intensity of 0.127 nJ/nm$^2$ per pulse at 100 kHz. The evaporation rate was controlled between 600 and 1200 events per second.

**2.4 Data Process and Visualization**

Data processing was carried out by the APyT Python modules (Eich, 2025). The process included mass spectrum calibration, chemical identification through analytical fitting of the spectra, and final three-dimensional reconstruction. Reconstruction was performed using a taper geometry, strictly following the shank angle and starting radius determined from SEM images of the prepared samples. The projection protocol in reconstruction followed the classical formalism of Bas et al. (Bas et al., 1995) and incorporated the boundary conditions established by Jeske et al. (Jeske & Schmitz, 2001). Visualization was performed in OVITO (Stukowski, 2010).

## 3. Results and Discussion

### 3.1 Establishing a Reproducible Preparation Window

While section 2 describes the practical implementation of the MicroCup-based preparation workflow, the purpose of this section is to rationalize its design and identify the key factors that enable its success. Organic molecular liquids impose constraints that are fundamentally different from aqueous systems and conventional solid materials: ion-beam exposure leads to pronounced deformation and chemical degradation, and ion irradiation of organic materials can also markedly reduce the effective milling rate because of the formation of modified surface layers (Tiddia et al., 2020). As a consequence, preparing large frozen organic droplets becomes far more time-consuming than preparing frozen water droplets with comparable sizes. In addition, unstable specimen geometries hinder reproducible field evaporation.

To address these challenges, the workflow in Section 2 was deliberately designed with three following strategies: minimizing beam-induced damage, improving time efficiency through controlled droplet volume, and achieving precise specimen morphology. In the following subsections, we demonstrate how each design element originates from these constraints and collectively defines a reproducible preparation window for APT analysis of organic molecular liquids.

#### 3.1.1 Minimization of Beam-Induced Damage during Imaging

In early tests, we observed that even one single ion-beam snapshot of 3 nA with a dwell time of 1 µs, caused pronounced morphological changes on the organic liquid samples. Despite being cryogenically frozen, the droplets exhibited progressive deformation: initial swelling, followed by local collapse and, in most cases, the formation of the internal voids (**Figure 2**b-d). Although the damaged sample could still be shaped into a specimen for measurement (**Figure 2**e), it is obvious that the internal structure of the droplet was severely distorted. By comparing the deposition thickness before swelling and after collapse, we found that a single ion-beam exposure removed nearly 2 µm of material from an initially 4 µm thick droplet. Moreover, beam-induced chemical degradation may also compromise molecular integrity, leading to inconsistent or non-representative fragmentation in mass spectra.

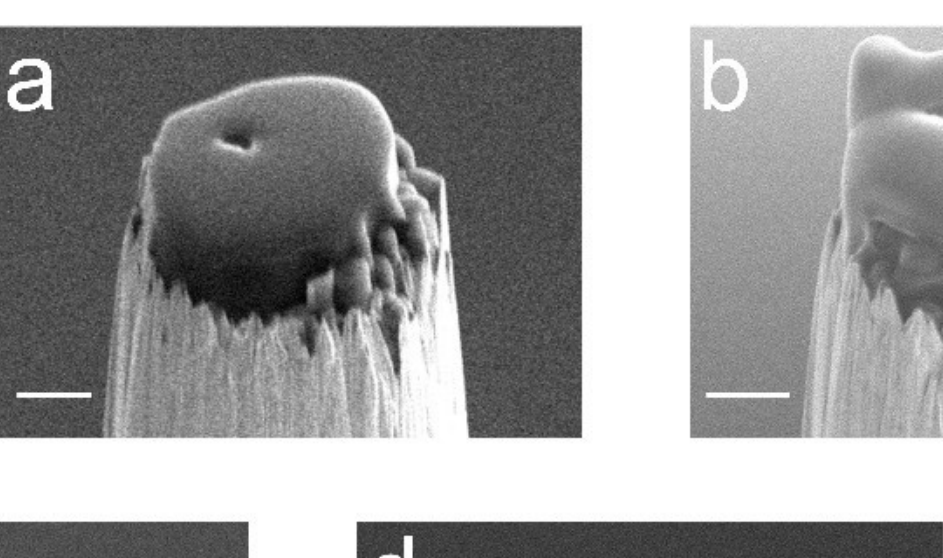

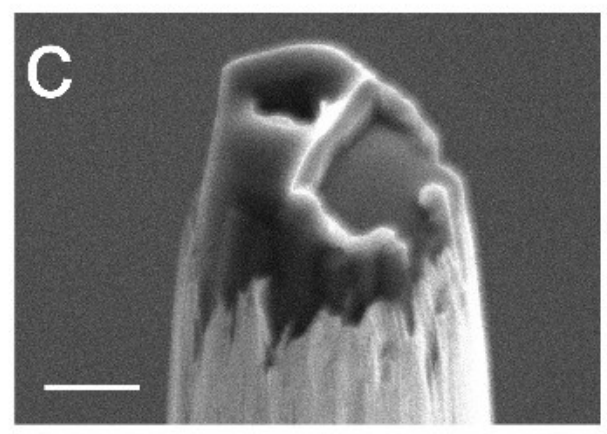

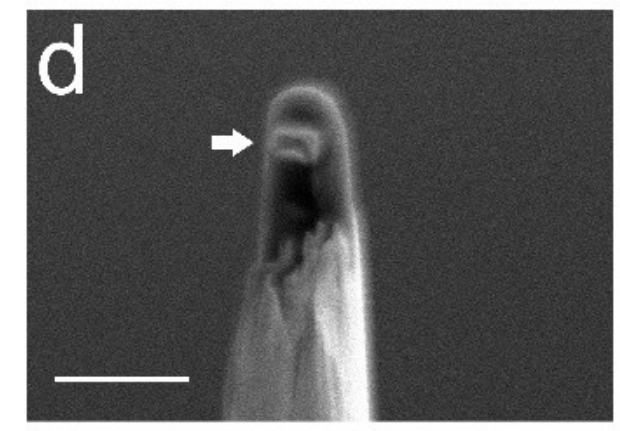

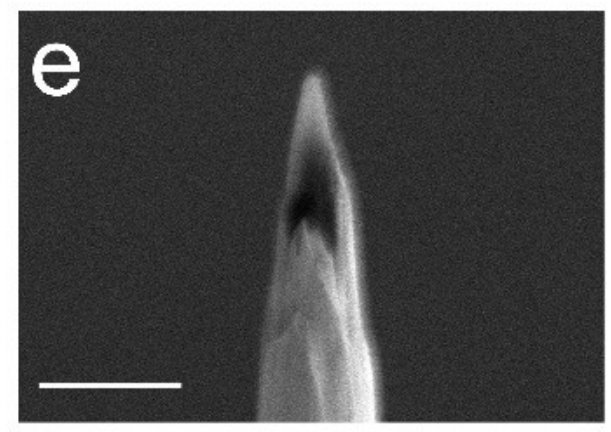

***Figure 2.*** *SEM images illustrating the beam-induced deformation of a frozen 8CB droplet after a single ion-beam exposure (30 kV, 3 nA, 1 μs dwell time) during specimen preparation. The deposited droplet on the W post (a) undergoes progressive swelling (b), local collapse (c), and internal void formation (d), highlighted by the arrow. The resulting specimen (e) fractured at the beginning of field evaporation, indicating substantial structural distortion and eventual pore formation caused by the ion beam. Scale bar: 2 μm.*

In conventional FIB workflows at room temperature, ion-beam-induced damage is commonly mitigated by depositing a sacrificial protective cap using a gas injection system (GIS) prior to imaging and milling. However, GIS-based deposition under cryogenic conditions remains difficult to control (Woods et al., 2023; Schreiber et al., 2018) and requires careful adjustment of the distance between the cold specimen surface and the GIS nozzle (Zhang et al., 2021). An alternative strategy is in situ metallic coating, which enables controlled deposition of thin metallic layers onto selected regions of flat sample surfaces directly inside the FIB (Schwarz et al., 2024). For frozen organic liquids, however, its implementation would require further optimization to minimize possible peak overlap from the coating material in mass spectrum and field-evaporation mismatch between the metal layer and the organic liquids (Larson et al., 2011; Oberdorfer & Schmitz, 2011). The heat impact from sputtering on the delicate surface structure of frozen organic liquids also remains to be clarified.

Therefore, ion-beam imaging above 0.3 nA was strictly prohibited throughout the entire preparation workflow within the area of interest. The milling process was predominantly monitored using low-dose electron beam imaging. When higher-current ion beam was used for milling, the beam exposure was only limited to the milling areas.

### 3.1.2 Time Efficiency

Before addressing the control of specimen morphology, another practical consideration must be discussed: time efficiency. Unlike frozen water, organic liquids interact chemically with Ga ions during FIB milling, forming a modified surface layer that significantly reduces the sputtering rate. As a consequence, the milling process for organic liquid becomes an extremely time-consuming process. For example, shaping a frozen droplet of n-tetradecane (**Figure 3**a) into a specimen requires over 10 hours (Meng et al., 2022), whereas an equivalently sized frozen water droplet can be milled in approximately 3-5 hours (Schwarz et al., 2020; Stender et al., 2022). More critically, this prolonged process must be carried out under strict beam exposure constraints to avoid damaging the sensitive organic material. This unique requirement forces the operator to remain highly focused throughout extended sessions. Because even small operational mistakes, such as minor misalignment or a misclick (**Figure 2**b) during imaging, can lead to sample failure.

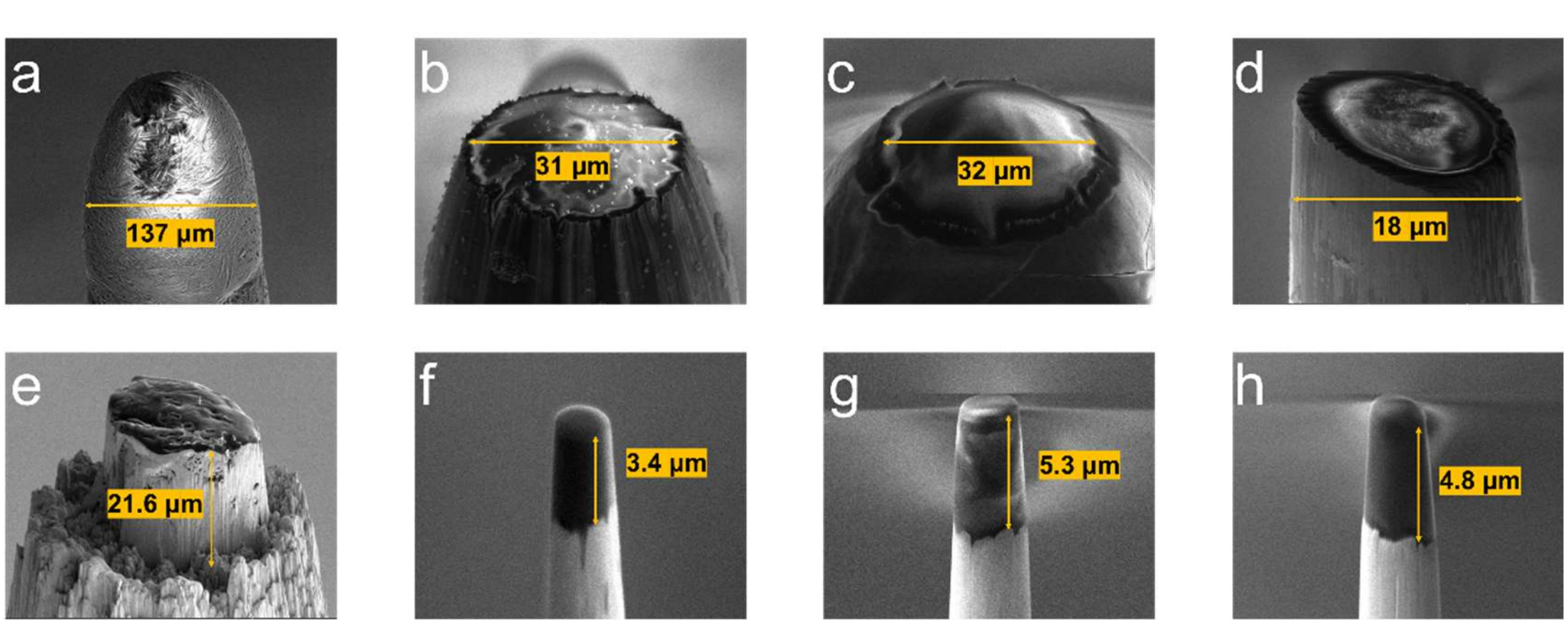

***Figure 3.*** *Comparison of the initial diameters and vertical thicknesses of the deposited organic molecular liquids. Panels (a,e), (b,f), (c,g), and (d,h) correspond to n-tetradecane deposited on cooled W post, 5CB, 8CB (deposited at pre-heated W post), and 8OCB (deposited into microcavity), respectively, showing the progressive reduction in droplet volume achieved through the MicroCup approach.*

To address this challenge, we identified minimizing the initial droplet volume as a critical factor in improving time efficiency. Reducing both the lateral diameter and the vertical thickness of the deposited liquid directly decreases the amount of frozen material that must be removed by cryo-FIB.

This process is summarized in **Figure 3**. In our earlier n-tetradecane study (Meng et al., 2022), the organic liquid was deposited onto a pre-cooled W post, producing frozen droplets with lateral dimensions exceeding 90 µm (**Figure 3**a) and vertical thicknesses above 20 µm (**Figure 3**e). Milling such a specimen required more than 10 h. In the subsequent 5CB/8CB experiments (Meng et al., 2026), the LC was deposited on a warm W post before quench freezing, which reduced the droplet diameter to approximately 35 µm (**Figure 3**b-c) and the vertical thickness to approximately 3 µm (**Figure 3**f). This reduced the average preparation time to below 7 h. With the MicroCup approach, the liquid volume was further confined by the microcavity, reducing the lateral dimension to approximately 10-15 µm (**Figure 3**d) and enabling full specimen fabrication in approximately 4 h, comparable to the preparation time reported for water-based specimens (Stender et al., 2022). Recent approaches for water-based specimens, including graphene encapsulation (Exertier et al., 2024) and nanoporous metallic substrate needles (Tegg et al., 2024), have also addressed the challenge of controlling the liquid volume. Together, these studies highlight that limiting the initial liquid volume is central to improving the preparation efficiency of liquid specimens for APT.

### 3.1.3 Morphology Control

The specimen morphology is determined primarily by three geometric parameters: the final specimen length, the shank angle, and the apex radius. While the apex radius can be precisely controlled by FIB annular milling parameters, we discuss below the control of final specimen length and the shank angle.

Based on limited beam exposure during the specimen preparation process, the final specimen length is determined by the initial deposition thickness of the droplet. In our very early 5CB/8CB experiments (Meng et al., 2026), both liquids were deposited at the same ambient temperature. Under these conditions, the deposition thickness reflected the inherent viscosity differences of the liquids: 5CB filled the cross-section of the cryo post cleanly, leaving sidewalls visible (**Figure 3**b), while the more viscous 8CB also coated side surfaces (**Figure 3**c), resulting in approximately 2 μm greater thickness (**Figure 3**f-g).

We found that this viscosity influence could be mitigated by heating the LC above the melting point before deposition. In the MicroCup approach, we added another layer of control: the deposition thickness was physically constrained by the cup's depth, which was precisely defined by the FIB milling process. A similar confinement principle was used by Li et al. for an ionic liquid in a Si microtip cavity (Li et al., 2019). The actual cup depth was verified by cross-sectional exposure during specimen preparation, as shown in **Figure S3.1**a. As shown in **Figure 3**g, 8OCB samples fabricated via MicroCup achieved a reproducible deposition thickness comparable to an 8CB specimen.

After deposition and quench freezing, the subsequent FIB shaping step can reduce the final specimen length if excessive beam exposure or over-milling occurs. Therefore, by minimizing beam exposure during shaping, the MicroCup-based workflow allowed reliable and reproducible control over the final specimen length, even for different classes of organic liquids (**Figure S3.1**b-c).

For non-wetting liquids (wetting angles $\vartheta > \pi/2$), the useful aspect ratios of depth $h$ to lateral radius $r$ of the cavity are likely restricted by the transition from the Wenzel to the Cassie-Baxter wetting regime (Marmur, 2003).

$$\frac{h}{r} < -\frac{1 + \cos\vartheta}{2\cos\vartheta} \quad (\text{with } \cos\vartheta < 0)$$

The shank angles were controlled during the final milling step at a reduced acceleration voltage (Perea et al., 2016; Langelier et al., 2017; Ji et al., 2020) of 8 kV. In earlier attempts, final shaping performed at 30 kV yielded low shank angles (**Figure 4**a), which always led to fracture during the beginning of the field evaporation process (in more than 30 trials). In contrast, milling at 8 kV consistently produced larger and mechanically more stable angles approximately to 20° (**Figure 4**b-c). Considering the relatively low heat conductivity of organic molecules, larger shank angles are known to support more stable field evaporation in organic systems (Proudian et al., 2019). The final apex radius was precisely controlled within the range between 100 and 200 nm.

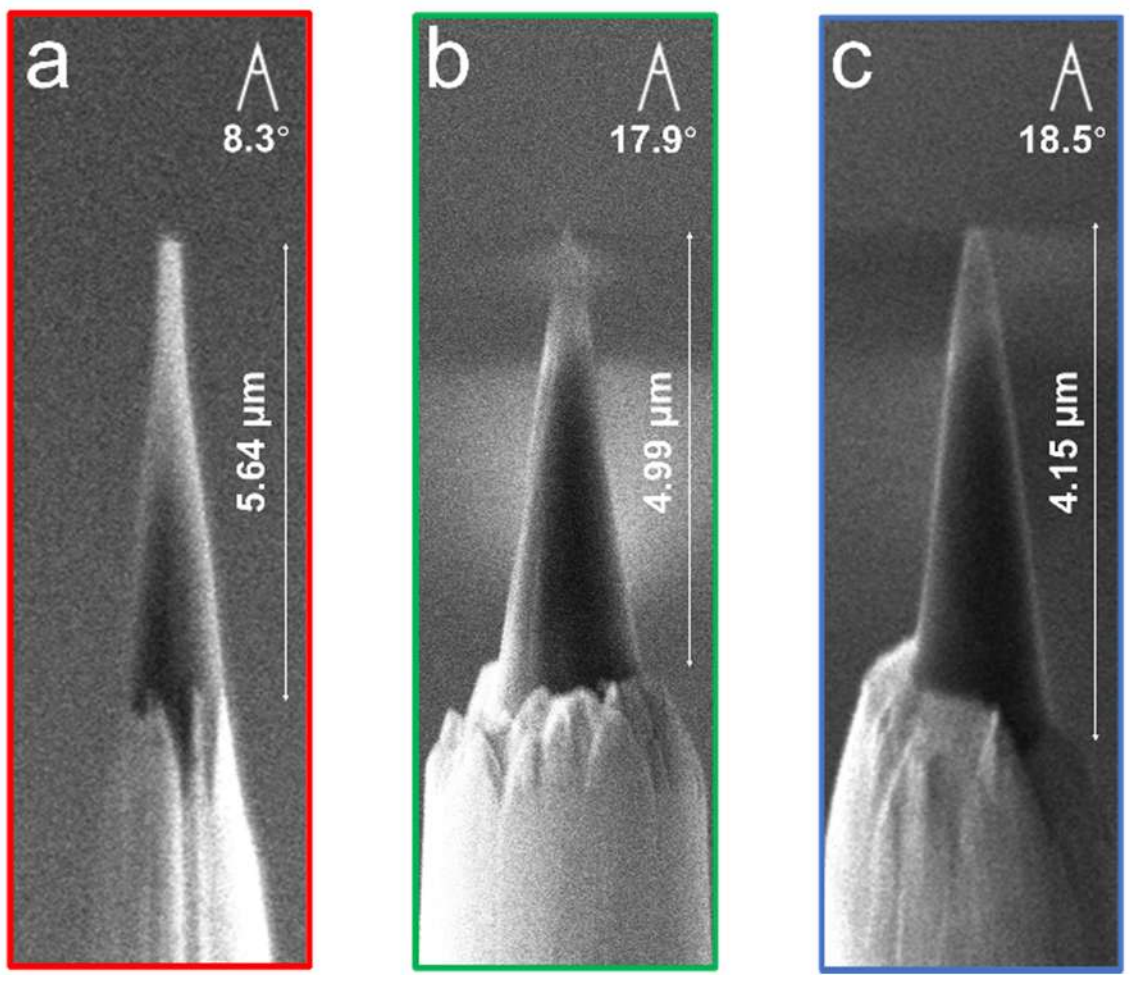

***Figure 4.*** *SEM images showing that larger shank angles in (b) 8CB and (c) 8OCB enable stable field evaporation, whereas the narrow-angle specimen in (a) 8CB fails.*

In summary, by combining the intrinsic fluidity of organic liquids with the high fabrication precision of FIB, the MicroCup approach enables a substantial improvement in preparation efficiency while providing precise control over specimen geometry. This methodology establishes a well-defined and reproducible geometric standard for organic liquid specimens, offering a practical reference for future studies in this field.

**3.2 Preservation of the Liquid State**

In cryogenic sample preparation and transfer, the specimen is inevitably analyzed in a frozen solid state, regardless of whether the starting material is a liquid or solid. The central question is therefore whether the frozen structure still faithfully reflects the original liquid configuration. This is only the case if crystallization is effectively suppressed during cooling, such that the liquid-derived molecular arrangement is preserved in a glassy, non-crystalline solid state.

If crystallization occurs, several fundamental limitations arise. First, the resulting structure no longer represents a liquid or mesophase but instead corresponds to a solid structure, precluding meaningful characterization of liquid-state configurations or solid-liquid interfaces. Second, crystallization of the solvent is inherently accompanied by solute rejection, leading to local solute enrichment and compositional deviations from the original liquid state (Brüggeller & Mayer, 1980). Third, the resulting concentration gradients and mechanical stresses may induce structural rearrangement, denaturation, or even fragmentation of dissolved molecular species (Zhang et al., 2025).

The formation of such a glassy, non-crystalline solid upon rapid cooling is referred to as vitrification (Wowk, 2010). From a kinetic perspective, a necessary condition for vitrification is that the effective cooling rate achieved in the experiment ($v_{exp}$) exceeds the critical cooling rate of the liquid under investigation ($R_c$). It is therefore necessary to estimate, at least on a semi-quantitative level, the range of cooling rates accessible under the MicroCup experimental conditions.

**3.2.1 Macroscopic Cooling Behavior**

At the macroscopic level, the entire assembly, including the MicroCup plus the metallic mounting components, requires approximately 90 s to cool from room temperature (298 K) to the temperature of liquid nitrogen (approximately 77 K). Here, the cooling time refers to the interval between immersion into the liquid nitrogen bath and the disappearance of the Leidenfrost effect. Under a linear cooling assumption, this corresponds to an average cooling rate of approximately 147 K/min (**Figure 5**a). Detailed calculation is provided in **Supplementary Section S3.1**. However, this value should only be seen as a conservative overall estimate. First, the cooling process is non-linear. Second, the focus of the study here is not the entire assembly, but the liquid inside the cup. Therefore, we introduce an upper-bound estimation specifically focused on the cup's cooling behavior.

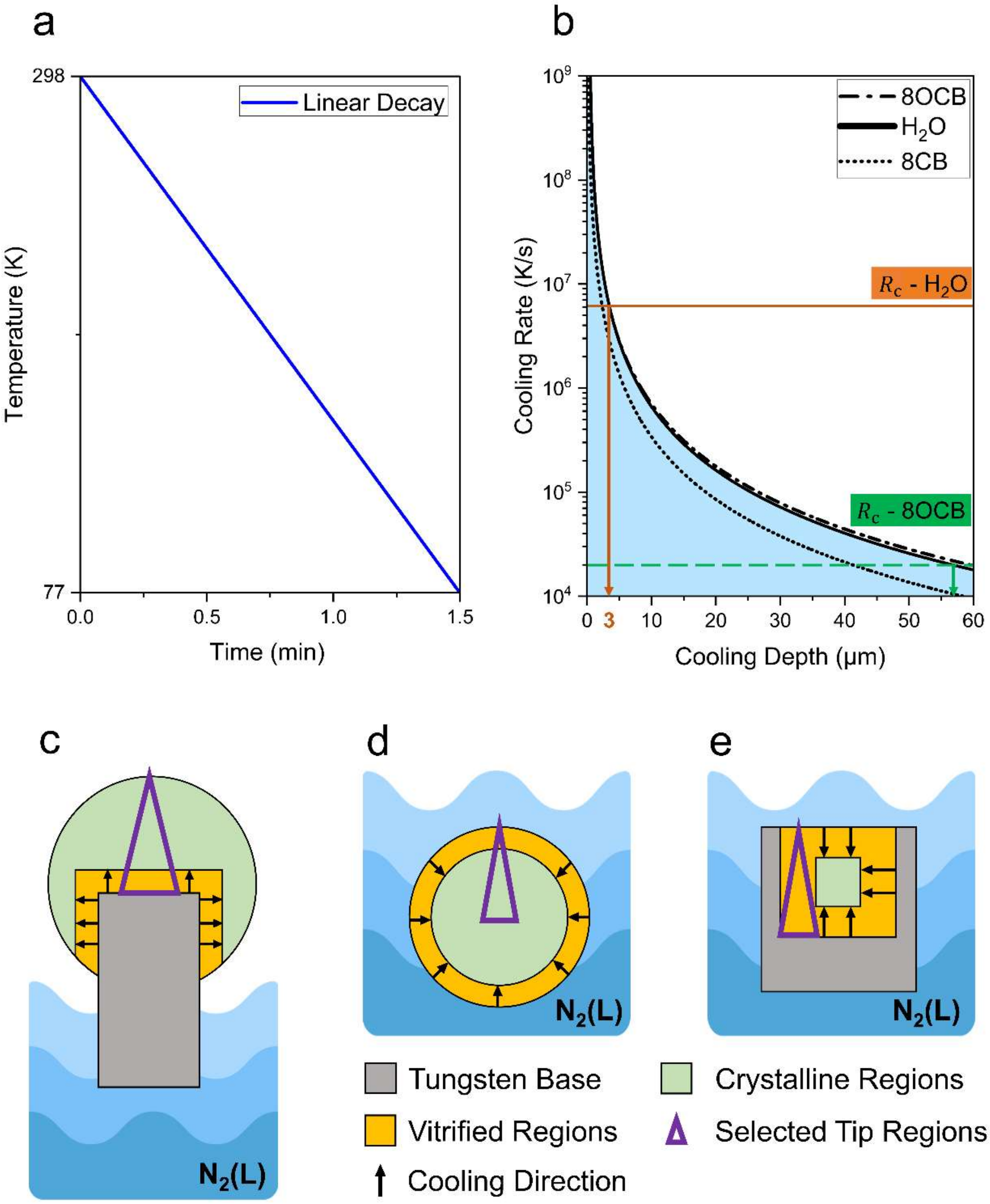

***Figure 5.*** *Cooling kinetics and vitrification geometry in cryogenic APT preparation. (a) Macroscopic cooling behavior of the MicroCup assembly during immersion in liquid nitrogen. The linear approximation yields an average cooling rate of ~147 K/min. (b)*

*Thermal diffusion-limited cooling rate as a function of cooling depth for different liquids, estimated using reported thermal diffusivities. Water (solid black line) is shown as a conservative benchmark, with its critical vitrification rate (~$6.4 \times 10^6$ K/s* (Mowry et al., 2025)*, orange line) defining a characteristic vitrification depth of approximately 3 μm. In contrast, 8OCB ($2 \times 10^4$ K/s* (Jiang et al., 2013)*, green dash-dotted line) exhibit markedly lower vitrification thresholds, corresponding to substantially larger vitrified depths under identical quenching conditions. (c–e) Schematic comparison of vitrified region locations in different cryo-APT preparation routes. (c) For pre-cooled wire deposition, heat extraction occurs primarily through the W substrate, confining vitrification to a thin interfacial layer at the bottom of the frozen droplet. (d) In cryo lift-out, rapid quenching produces a thin vitrified shell near the droplet surface, which may coincide with the final specimen apex but requires special cares to material removal during ion-beam shaping. (e) In the MicroCup geometry, the reduced deposition volume and lateral heat extraction from the cup walls bring vitrified regions from multiple directions into closer proximity with the final APT specimen, thereby maximizing the probability that the analyzed volume originates from a vitrified, liquid-derived region.*

### 3.2.2 Microscopic Thermal Diffusion Approximation

At the specimen scale, heat extraction is dominated by thermal diffusion from the liquid nitrogen into the frozen droplet. To establish an upper bound for the achievable quench rates, we adopt a diffusion-limited approximation that neglects the Leidenfrost effect and assumes efficient heat transport through the W substrate. Within this framework, the attainable cooling rate decreases rapidly with increasing distance from the heat sink, defining a characteristic vitrification depth rather than a uniform cooling profile (**Figure 5**b). A detailed derivation of the diffusion-limited vitrification depth is provided in the **Supplementary Section S3.2**.

To provide a conservative reference for the cooling-rate magnitude, water is considered as a benchmark system for which critical cooling rates have been extensively reported in the literature (Brüggeller & Mayer, 1980; Dubochet & Lepault, 1984; Tanaka & Kimura, 2019; Mowry et al., 2025). Here, a conservative value of $6.4 \times 10^6$ K/s is adopted (Mowry et al., 2025), which corresponds to an amorphous layer thickness of approximately 3 μm (**Figure 5**b). Beyond this depth, the experimental cooling rate is expected to fall below that required to preserve liquid-like structure in water. Now let us discuss where is this vitrified layer located in different specimen preparation procedures.

For preparation routes in which the W wire is pre-cooled prior to droplet deposition, cooling proceeds predominantly via heat transfer from the cold W substrate into the liquid (**Figure 5**c). This implies that the vitrified layer is located at the bottom of the frozen droplet. Given that the initial thickness of deposited water droplets (>20 μm) is typically much larger than 3 μm, the majority of the droplet is expected to crystallize during quenching. As a result, the APT specimen formed from such a droplet, particularly the apex region probed during analysis, is likely to originate predominantly from crystallized water rather than from the narrow vitrified interfacial layer.

In the case of cryo lift-out, a large droplet (diameter >30 μm) is first rapidly quenched in liquid nitrogen. Under ideal thermal conduction assumptions, this 3 μm thick vitrified layer should be located at the outer shell of the frozen droplet, which spatially coincides with the apex region ultimately prepared for APT analysis (**Figure 5**d).

In principle, this geometry allows access to liquid-derived amorphous water. However, given the limited thickness of this vitrified layer, its preservation during subsequent ion-beam milling requires particular care, as extensive material removal, vacuum exposure, or beam-induced heating could erode or transform this near-surface region before measurement.

By contrast, the MicroCup geometry increases the likelihood of preserving vitrified liquid structures within the prepared APT specimen. Owing to the substantially reduced initial deposition volume, the characteristic vitrified regions originating from the cup opening and the W base are brought into closer proximity along the axial direction. By choosing a cup size less than the vitrification depth, it is likely that a homogeneously amorphous sample may be produced. In addition, site-specific specimen preparation can be selectively performed near the cup walls, where lateral heat extraction further enhances the spatial overlap between vitrified regions and the final APT specimen (**Figure 5**e). Together, these geometric factors maximize the probability that the analyzed specimen volume originates from a vitrified, liquid-derived region, rather than from crystallized material.

In contrast to water, many low-molecular-weight organic liquids exhibit significantly reduced crystallization kinetics and can undergo vitrification at comparatively modest cooling rates (Tsuji et al., 1971; Wedler et al., 1991; Dierking, 2008; Massalska-Arodź, 2024). For instance, 5CB has been reported to remain vitrified at cooling rates exceeding approximately 5 K/min (Mansaré et al., 2002), which is already well below the linear macroscopic cooling rate of the MicroCup assembly (~70 K/min). For 8CB and 8OCB, direct measurements of critical cooling rates are scarce. However, experimental studies have demonstrated that 8OCB can be vitrified at cooling rates above $2\times10^4$ K/s (Jiang et al., 2013). Within the diffusion-limited framework established above, this rate corresponds to a characteristic cooling depth exceeding 50 μm (**Figure 5**b), which is way beyond the geometric dimensions of the MicroCup configuration.

In the following section, we demonstrate, using 8CB and 8OCB as model systems, how this approach effectively controls the field-evaporation behavior of small organic molecules while preserving their intrinsic chemical composition and structural information.

### 3.3 Representative Example: 8CB vs 8OCB

#### 3.3.1 Field Evaporation Behavior

Before discussing the field evaporation behavior of organic molecules, it is necessary to verify that the specimen-preparation workflow can preserve the molecular structure. This is particularly important for beam-sensitive organic molecules, where ion- or electron-beam exposure during specimen preparation may already fragment molecules before APT analysis.

We first focused on phase-homogeneous smectic-like regions, where intact molecular ions of both 8CB and 8OCB were reproducibly detected across different specimens (**Figure S3.1**d). The intact-molecule fractions were then quantified from fitted and integrated peak contributions, followed by carbon-normalized analysis. The detailed calculation procedure and the corresponding results are provided in **Supplementary Section S5**. Under these conditions, both 8CB and 8OCB evaporated predominantly as intact molecular ions, with carbon-normalized intact-molecule fractions above 90% for 8CB and above 70% for 8OCB (**Figure 6**f). Such high intact molecular fraction is consistent with minimal preparation-induced damage (Rivas et al., 2020), although the apparent intact-molecule yield in APT can also depend on experimental conditions such different fields and pulsing modes (Perea et al., 2017). This intact evaporation effectively reduces the ambiguity commonly associated with small organic fragments in interpretation (Schwarz et al., 2021; Zhang et al., 2025). These molecular peaks also provide clear spectral references for comparing 8CB and 8OCB. The dominant molecular ion in 8CB spectrum is the singly charged $C_{21}H_{25}N^{+}$ near m/z= 291 u, accompanied by its dehydrogenated forms ($C_{21}H_{23}N^{+}$ and $C_{21}H_{21}N^{+}$) and a minor contribution from the doubly charged $C_{21}H_{26}N^{++}$ near m/z= 146 u. In 8OCB, the same intact molecular signals are observed, but respective peaks shift by approximately 16 u due to the additional oxygen atom (compare yellow regions in **Figure 6**a-b).

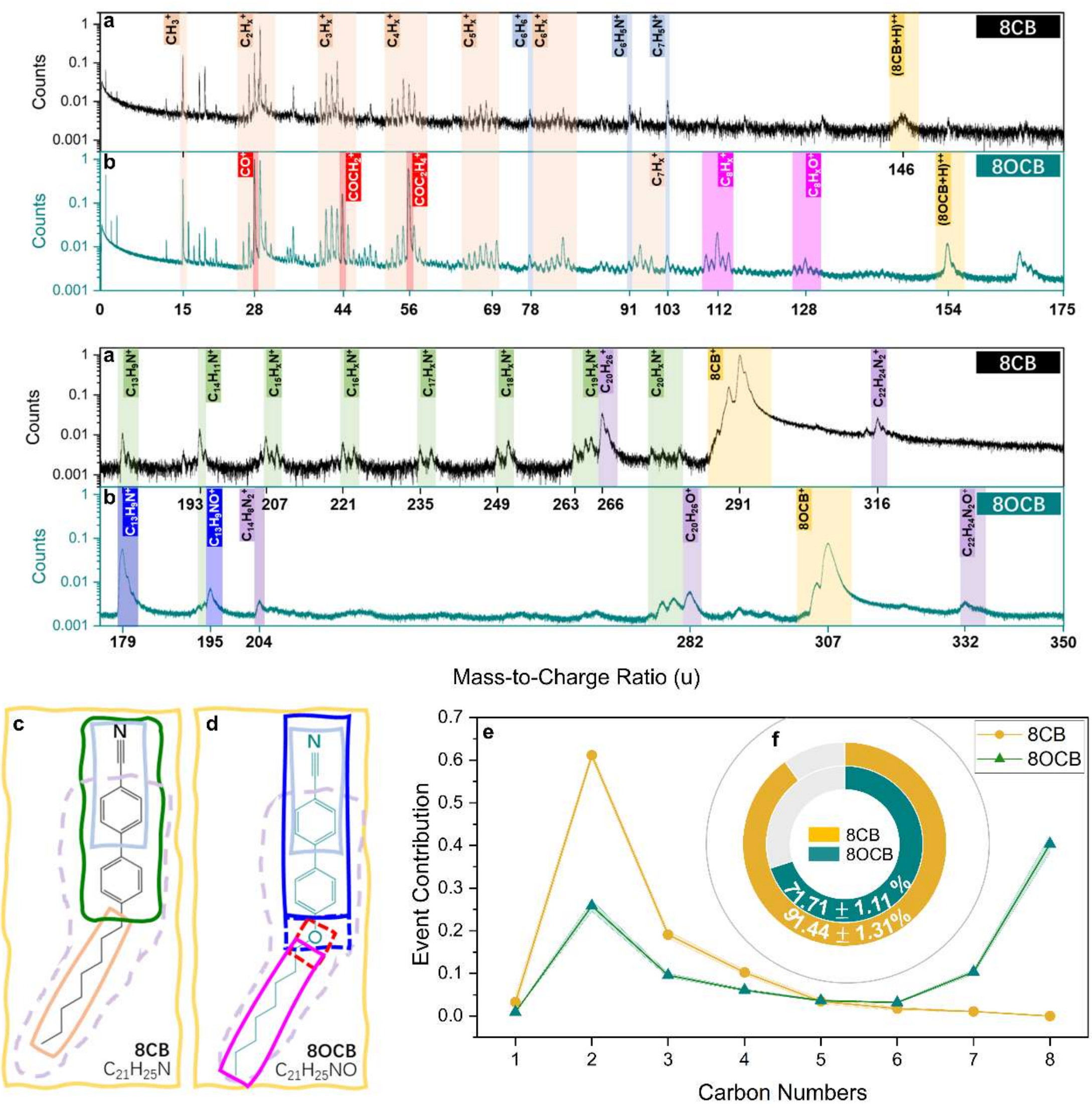

***Figure 6.*** *Direct comparison of the normalized mass spectra and fragmentation behavior of 8CB and 8OCB. (a) pure 8CB and (b) pure 8OCB within 0–350 u. (c,d) Schematics of*

*the corresponding fragmentation pathways. (e) Event contributions of C1–C8 alkyl fragments in 8CB (yellow) and 8OCB (green). (f) fraction of intact molecules detected in smectic phases of both systems.*

Additional evidence for preserved molecular integrity is provided by the minimized Ga-related contamination. First, the event contribution of the peak near 69 u is below 0.2% in both datasets (**Table S8.1**). Therefore, regardless of whether this signal is assigned to $^{69}Ga^+$ or to the organic fragment $C_5H_9^+$, its low abundance indicates that the MicroCup workflow effectively minimizes Ga contamination. Second, the spatial distribution of this signal in both 8CB and 8OCB datasets does not show the characteristic enrichment at the beginning of the analysis that would be expected for surface or preparation-induced Ga implantation (**Figure S8.1**). Furthermore, the measured peak positions at 69.07 u in 8CB and 69.04 u in 8OCB are clearly offset from the expected position of $^{69}Ga^+$ at 68.93 u (**Table S2-3**). This systematic offset more strongly supports an organic-fragment assignment. Together, these observations support that the restricted ion-beam exposure during preparation effectively minimizes Ga contamination and is consistent with preserved molecular integrity.

With this reliable molecular baseline, we can now examine the fragmentation behavior in more detail. In both systems, the detected fragment ions fall into four reproducible categories, which provide a clear framework for comparing their fragmentation pathways.

Alkyl fragments, mostly derived from the terminal alkyl chain, consist of ions ranging from $C_1H_x^+$ to $C_8H_x^+$, which means the combination of one to eight carbon atoms with varying numbers of hydrogen atoms. Monophenyl derivatives, detected primarily as $C_6H_6^+$, $C_6H_5N^+$ and $C_6H_5CN^+$, reflect species in which a phenyl ring is bonded to a nitrogen atom or a cyano group. Cyanobiphenyl residues exhibit a characteristic mass series beginning near m/z = 179 u with the cyanobiphenyl core $H_4C_6C_6H_5CN^+$ (labeled as $C_{13}H_9N^+$), increasing in regular 14 u intervals (corresponding to successive $CH_2$ units) and extending up to $C_{20}H_xN^+$. Finally, a pair of ions identified as cyano-hydrogen substitution pairs are located approximately ±25 u relative to the molecular ion peak in both 8CB and 8OCB systems. The peak $C_{20}H_{26}^+$ in 8CB or $C_{20}H_{26}O^+$ in 8OCB systems represents the molecule experience a cleavage of the cyano group then followed by addition of a hydrogen. In the meanwhile, the cleaved cyano group replaced a hydrogen of another molecules, forming $C_{22}H_{24}N_2^+$ in 8CB or $C_{22}H_{24}N_2O^+$ in 8OCB. For clarity, we assign different colors to these four ion types in **Figure 6**a and **6**b: alkyl fragments (orange), monophenyl derivatives (blue), cyanobiphenyl residues (green), and cyano-hydrogen substitution pairs (purple). Schematics of the fragmentation behavior of both molecules are sketched in **Figure 6**c and **6**d respectively.

Despite this overall similarity, the addition of a single oxygen atom leads to three systematic differences between 8CB and 8OCB.

First, the cyanobiphenyl residues show markedly different truncation behavior (green labels in **Figure 6**a-b). In 8CB, the fragment series evaporates in regular 14 u intervals from $C_{13}H_9N^+$ up to $C_{20}H_xN^+$, whereas in 8OCB only the shortest residue $C_{13}H_9N^+$ is detected (highlighted in blue in **Figure 6**b). This indicates that the additional oxygen atom introduces a preferential rupture site along the molecular backbone and selectively preserved the shortest cyanobiphenyl core.

Second, this rupture near the oxygen atom influences the evaporation of the alkyl chain, leading to opposite fragment contribution trends between the two systems (**Figure 6**e). In 8CB, the event contribution of alkyl fragments decreases progressively from C2 to C8, reflecting the instability of longer chains under high electric fields. In contrast, 8OCB displays an increasing contribution from C6 to C8 fragments, which reveals that oxygen-induced cleavage promotes the integrity of the remaining alkyl chain (highlighted in magenta in **Figure 6**b).

Third, the resulting evaporation products include distinct oxygen-containing fragments in 8OCB, with pronounced peaks near m/z = 28, 44, and 56 u. Because peaks in soft molecular systems can contain overlapping contributions from multiple candidate ions (Schwarz et al., 2021), the following discussion refers to the most plausible dominant contributions rather than unique peak identities. To evaluate these assignments, the spectra were constructed using a 0.01 u bin width for fine-bin visualization and fitting (Eich, 2025), not as a measure of the instrumental mass resolving power. The dominant assignments were then assessed using three complementary criteria: comparison of the relative peak intensities among 5CB, 8CB, and 8OCB (**Figure S7.1**c); evaluation of small but systematic peak shifts near 28, 44, and 56 u among these systems (**Figure S7.1**d); and fitting of the experimental peaks with theoretical curves (**Figure S7.1**e). The detailed discussion is provided in **Supplementary Section S7**. This combined analysis supports the assignment of $CO^+$, $COCH_4^+$, and $COC_2H_4^+$ as the dominant contributors to the peaks near 28, 44, and 56 u, respectively (highlighted in red in **Figure 6**b). The emergence of these ether-related fragments indicates that the oxygen site not only defines a preferential cleavage region but also redirects the chemical pathway toward oxygen-containing ionization products.

These observations can be rationalized by the electronic effect introduced by the oxygen atom. The high electronegativity of oxygen polarizes the adjacent C-C bonds, which lowers the bond dissociation energy at the substitution site and defines a preferential cleavage point along the alkyl chain. Under the intense local electric fields during APT, this polarization is further amplified, making the oxygen-adjacent bond the dominant rupture site. This interpretation is consistent with earlier experimental and theoretical studies on the fragmentation of organic molecules under high electric field conditions. Early scanning atom probe experiments showed that the dissociation of organic molecules is highly characteristic and reflects the underlying molecular binding state rather than random fragmentation (Nishikawa et al., 2006). Theoretical work on conjugated polymers further suggested that strong electrostatic fields can induce charge redistribution and weaken specific intramolecular bonds (Wang et al., 2006).

Interestingly, in 8OCB both the remaining alkyl chain and the cyanobiphenyl core tend to evaporate in a more intact and directional manner (highlighted in **Figure 6**d). Such behavior may ultimately enable the use of fragmentation patterns in APT to probe orientational order in future molecular systems.

Crucially, these systematic and structurally linked evaporation pathways only become apparent under stable and geometry-controlled evaporation conditions, highlighting the importance of specimen morphology control for mechanistic interpretation.

Taken together, these results demonstrate that the incorporation of a single oxygen atom at a chemically strategic location is sufficient to fundamentally alter the molecular fragmentation pathway. This highlights the critical role of local electronic structure in governing field evaporation behavior and opens new opportunities for peak identification and tuning molecular stability and ionization routes through a rational chemical design.

### 3.3.2 Structural Identification of Solid-Liquid Interfaces

While the mass spectra obtained from 8CB and 8OCB exhibit clear and chemically meaningful signals, they do not necessarily translate directly into reliable spatial reconstructions (**Figure S9.1**). This is particularly true for liquid systems, whose lack of translational and orientational order makes it currently impossible to validate spatial reconstruction using APT. To address this, we intentionally introduced partial crystallization via thermal treatment within the MicroCup, thus facilitating the in-situ generation of tailored solid-liquid interfaces. This approach allows us to evaluate the reliability and applicability of spatial reconstruction in the soft molecular systems.

To perform the heat treatment, both 8CB and 8OCB were filled into the MicroCups at their respective isotropic temperatures, followed by cooling below their crystallization thresholds and holding at 283 K for 10 minutes in 8CB or at 295 K for 30 minutes in 8OCB (**Figure S12.1**). These durations are sufficient to initiate nucleation but intentionally too short to permit substantial grain growth. This controlled partial crystallization yields coexisting solid and liquid regions rather than fully developed crystals, thereby generating well-defined solid-liquid interfaces within the MicroCup cavity. Finally, the samples were rapidly quenched in liquid nitrogen to preserve these transient structures.

In a heat treated 8CB sample, we identified two regions showing pronounced molecular crystallographic periodicity, mostly contributed by intact 8CB molecules (**Figure 7**a). Such clear one-dimensional translational order could be interpreted in two ways: they might indicate a smectic phase or they could arise from small crystalline nuclei. However, given the limited grain size, spatial distribution map data are insufficient to determine whether an additional in-plane translational periodicity is present.

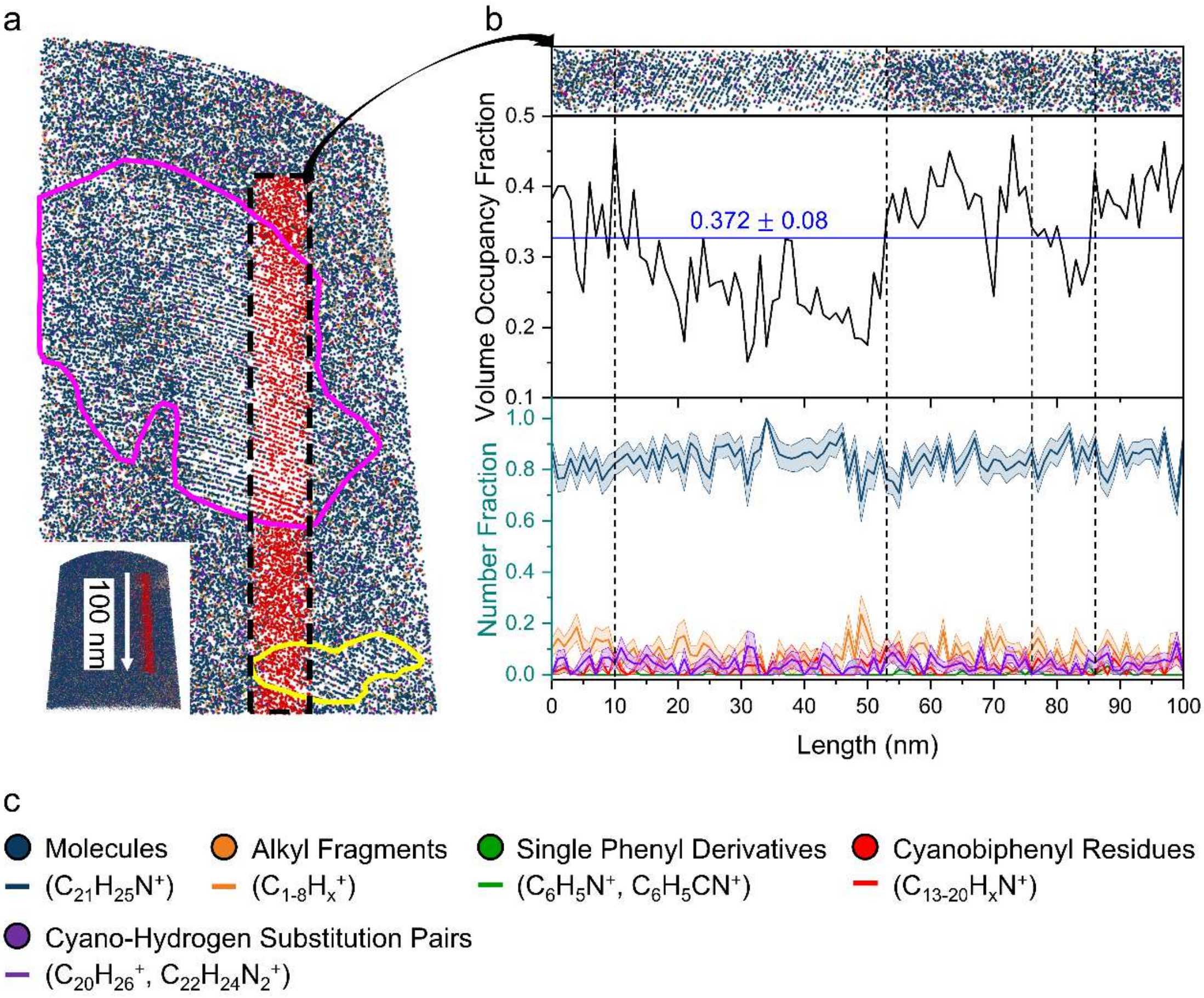

***Figure 7.*** *Direct visualization of 8CB crystalline nuclei embedded within a smectic matrix. (a) Two crystalline regions exhibiting distinct layered structures are highlighted in magenta and yellow. The red cylinder (5 nm radius, 100 nm length) indicates the region used for one-dimensional volume occupancy fraction and number fraction analysis. The bottom-left inset shows the full reconstructed specimen. (b) One-dimensional volume occupancy fraction and number fraction profile along the cylindrical region. (c) Color coding of the key ion species represented in panels (a) and (b).*

Because the evaporation fields of 5CB, 8CB, and 8OCB are unknown and may vary between smectic-like and crystalline regions, reconstruction was performed using a geometry-constrained model rather than a voltage-based radius evolution. Based on the precise specimen geometry control (**Figure 4**b), the taper angle chosen for reconstruction is constrained to a narrow range between 16° and 20°. Within this window, the measured interlayer spacing remains from 0.6 to 0.8 nm, which is far below the smectic layer spacing of 3.173 nm (Safinya et al., 1993). Reproducing 3.173 nm would require an unrealistically small reconstruction angle (<0.5°), which is incompatible with the observed specimen geometry. Thus, this supports the assignment to a crystalline phase. Additional supporting evidence including the local curvature relationship, pole-like evaporation feature, SDM analysis, and artefact comparison are provided in **Supplementary Section S10**.

Furthermore, we quantified the variation of volume filling (event density) between the crystalline domain and adjacent regions. Here, the volume occupancy fraction (VOF) represents the accumulated reconstructed volume fraction contributed by each ion group and is used as the reconstructed event density proxy along the analysis cylinder. As shown in **Figure 7**b, the density difference reached nearly 100%, indicating a substantial contrast in evaporation threshold between the two phases. This phenomenon is likely attributable to the weaker intermolecular interactions in the smectic phase, due to its larger layer spacing, which enhances local magnification (Miller & Hetherington, 1991) of the crystalline regions during co-evaporation. The observed density contrast thus provides indirect yet compelling evidence of solid-liquid coexistence.

Notably, this study provides a rare direct visualization of crystalline domains embedded in soft organic matter via APT and a glimpse into a solid-liquid interface in an organic molecular system. These findings recommend the MicroCup strategy in mitigating beam-induced damage while preserving thermodynamically unstable molecular structures. Moreover, they showcase the potential of APT in probing local ordering and interfacial behavior in molecular liquids.

In heat-treated 8OCB samples, similarly well-defined heterogeneity between high-density (beige) and low-density (green) domains was observed (**Figure 8**a). A one-dimensional profile over a 70 nm axial range (**Figure 8**b) revealed that the high-density beige region predominantly contains intact molecules and cyanobiphenyl residue fragments, whereas the low-density green domain is enriched in smaller alkyl and oxygen-containing ether fragments. The corresponding number-fraction profile, which provide the detailed comparison among different fragment types, are provided in **Figure S11.1**b. This spatial contrast is consistent with the mass spectrometric observation that the ether linkage weakens molecular stability in 8OCB.

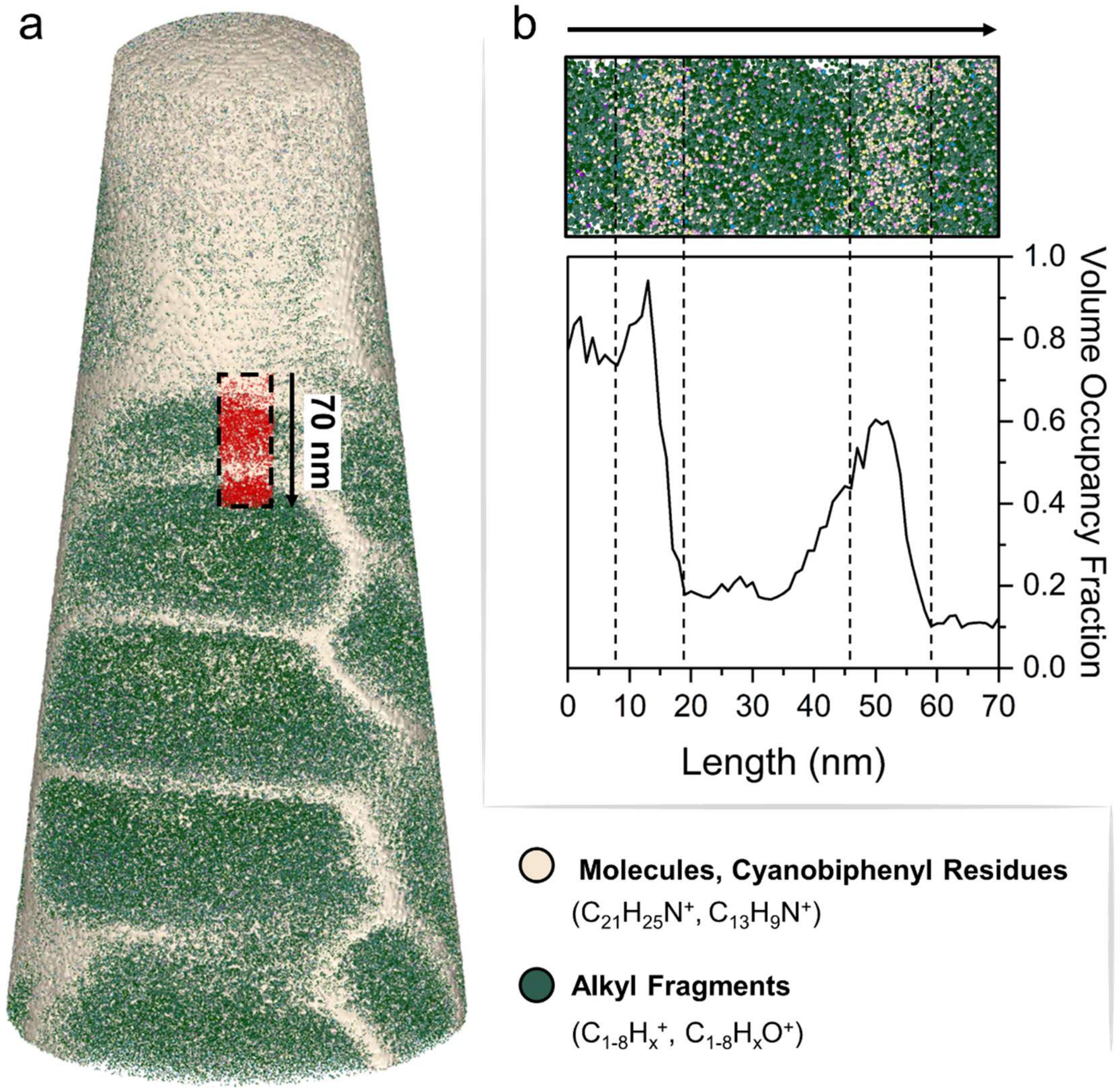

***Figure 8.*** *Visualization of solid-liquid interfaces in pure 8OCB. (a) Reconstructed 8OCB specimen showing crystalline domains embedded in surrounding smectic network. The beige region corresponds to intact molecules and major molecular fragments, while the green region is dominated by small alkyl fragments, indicating compositional and structural differentiation. (b) One-dimensional volume occupancy fraction profile across the solid-liquid interface highlighted in (a). The analysis was performed over a cylindrical region (10 nm radius, 70 nm length).*

We hypothesize that the low-density green domain corresponds to crystalline regions. However, due to molecular fragmentation during evaporation, direct observation of lattice planes in 8OCB is not possible. Nevertheless, we propose that tighter molecular packing and stronger interactions result in higher evaporation fields and local magnification, thereby yielding reduced apparent density in reconstruction. Typically, higher field strengths also lead to smaller fragments in evaporation. Conversely, smectic-like regions, with more disordered packing and weaker interactions, evaporate at lower fields, mostly in form of intact chains and leading to higher apparent densities. This finding also aligns with previous reports noting the preferential evaporation of intact organic molecules in liquid states

(Schwarz et al., 2021).

Although we have constructed a multi-dimensional body of evidence, including layered morphology, reconstruction spacing, specimen geometry, and spatial density, to support the coexistence of crystalline and smectic phases in 8CB and 8OCB, further experimental validation is still recommended. Future correlative cryogenic transmission electron microscopy could provide complementary structural information and help assess the accuracy of APT reconstructions once suitable workflows for preserving and imaging thermotropic LC systems are established. Current limitations in detector efficiency (<50%) and measurement temperature (60 K) restrict the overall signal quality and spatial resolution. We anticipate that the adoption of next-generation instrumentation, offering higher detection efficiency (~80%) and lower measurement temperatures (~40 K), will significantly enhance the quantitative capabilities of APT in soft matter systems.

In summary, this structural and evaporation analysis confirms the viability of our MicroCup-based approach in probing the complex phase behavior of organic molecular liquids. It further demonstrates the potential of APT to distinguish phase-specific characteristics, identify interfacial architectures, and elucidate field evaporation mechanisms. The geometric control and beam mitigation afforded by the MicroCup technique establish a solid methodological foundation for future APT studies of soft matter, interfacial systems, and functional organic materials.

## 4. Conclusions

This work establishes a feasible and systematic pathway for APT to organic molecular liquids. The core advancement lies in confining the liquid within a MicroCup cavity prior to solidification, enabling precise control over specimen geometry and providing the conditions necessary for reproducible field evaporation. A practical geometric parameter window is identified with a final specimen length of 4-5 μm, an apex radius close to 100-200 nm, and a taper angle of approximately 20°, which provides a reference morphology for future studies on organic materials in APT. The reduced starting volume inherent to the MicroCup design further improves preparation throughput.

Spectral interpretability is improved by minimizing ion beam exposure prevents structural degradation and suppresses excessive fragmentation, which resulted in spectra dominated by intact molecular or large fragment ions with less controversy. Ambiguous small-fragment peaks were clarified by comparing different systems and checking their exact mass positions, which allowed us to confirm that the oxygen atom in 8OCB promotes selective bond cleavage through local bond polarization. Structurally, the slower cooling kinetics of organic molecules permit controlled partial crystallization within the MicroCup, producing in situ solid-liquid coexistence. One-dimensional volume occupancy profiles reveal the spatial position of the interface and demonstrate phase-dependent evaporation behavior: 8CB remains largely invariant, whereas 8OCB exhibits distinct smectic-like and crystalline responses.

Overall, this approach provides a methodological foundation for extending APT to molecular liquids, enabling controlled morphology, interpretable mass spectra, and spatial analysis of phase coexistence. It opens a pathway toward systematic studies of interfacial physics, molecular organization, and liquid-phase behavior in functional soft materials.

## 5. Outlook

Beyond its technical robustness, the MicroCup strategy shows broad applicability. It is inherently suited for organic liquids and soft solids that tolerate thermal melting without degrading molecular architecture, such as thermotropic LCs, ionic liquids (**Figure S3.1**c), or molecular glasses. It can also accommodate solute-in-solvent systems by filling the MicroCup with an appropriate solution and removing the solvent post-deposition. Although not explored here, the approach may be extendable to water-based systems by tuning MicroCup geometry to prevent ice expansion and surface fracture during vitrification.

In addition, the MicroCup approach also offers distinct advantages over cryo-liftout methods. First, the feasibility of applying high-current ion beams (e.g., 3 nA) to organic droplets is questionable, our preliminary tests reveal severe deformation and collapse under such conditions, raising concerns about the integrity of molecular systems during lift-out procedures. In contrast, the MicroCup method enables minimal beam exposure by confining beam interaction to regions away from the liquid itself. Second, the typical challenge of specimen attachment in cryo-liftout (Schreiber et al., 2018; Woods et al., 2023) is inherently resolved, as liquid deposition within the MicroCup achieves natural anchoring without the need for additional manipulation (**Figure S13.2**). Third, the method substantially lowers instrumental barriers: instead of requiring advanced cryo-EasyLift system, it only involves drilling a shallow cavity into an electropolished wire specimen, a task accessible to most FIB platforms.

Moreover, the MicroCup method enables geometric flexibility beyond simple volume control. The cavity shape can be customized to support specific molecular orientations; alignment layers, such as grooves, may be patterned on the inner walls akin to twisted-nematic LCD fabrication (Kawata et al., 1994); and external fields may be applied during filling to induce orientational order. These extensions open opportunities for controlling molecular conformation inside the probe volume, thereby allowing deeper insights into how nanoscale packing, dipole alignment, or conformational heterogeneity influence field evaporation behavior. This control could also mitigate irregular fragmentation and advance the structural characterization of complex organic materials by APT.

## 6. Acknowledgements

We gratefully acknowledge financial support from the Deutsche Forschungsgemeinschaft (DFG, German Research Foundation) under Project-ID 358283783 - SFB 1333.

# Supplementary Information

This document provides supplementary figures, tables, and methods supporting the results discussed in the main manuscript.

**Title:**

MicroCup: A Cryogenic Specimen Preparation Strategy for Atom Probe Tomography of Organic Molecular Liquids

**Authors:**

Kuan Meng[1], Florian Groll[1], Sebastian Eich[1], Guido Schmitz[1]

**Affiliations:**

1) University of Stuttgart, Institute for Materials Science, Germany

**Corresponding Author Email Address:**

Kuan.Meng@imw.uni-stuttgart.de

# Menu

## S1. Requirement of an exposed single-apex geometry

In the MicroCup-based workflow, the cavity is used as an intermediate containment structure and not as the final analysis geometry. After the liquid is confined and solidified, the filled MicroCup must be shaped into a needle-shaped specimen with an exposed single apex.

This requirement was established from early preparation trials. **Figure S1.1**a-b show two early 5CB specimen in which the organic apex appeared well shaped, but a neighboring tungsten (W) base remained close to the intended analysis region. Despite the apparently suitable specimen geometry, both specimens did not produce a stable molecular signal during APT measurement. This failure indicates that nearby metallic supports could perturb the local electric field and reduce the effective field at the organic apex.

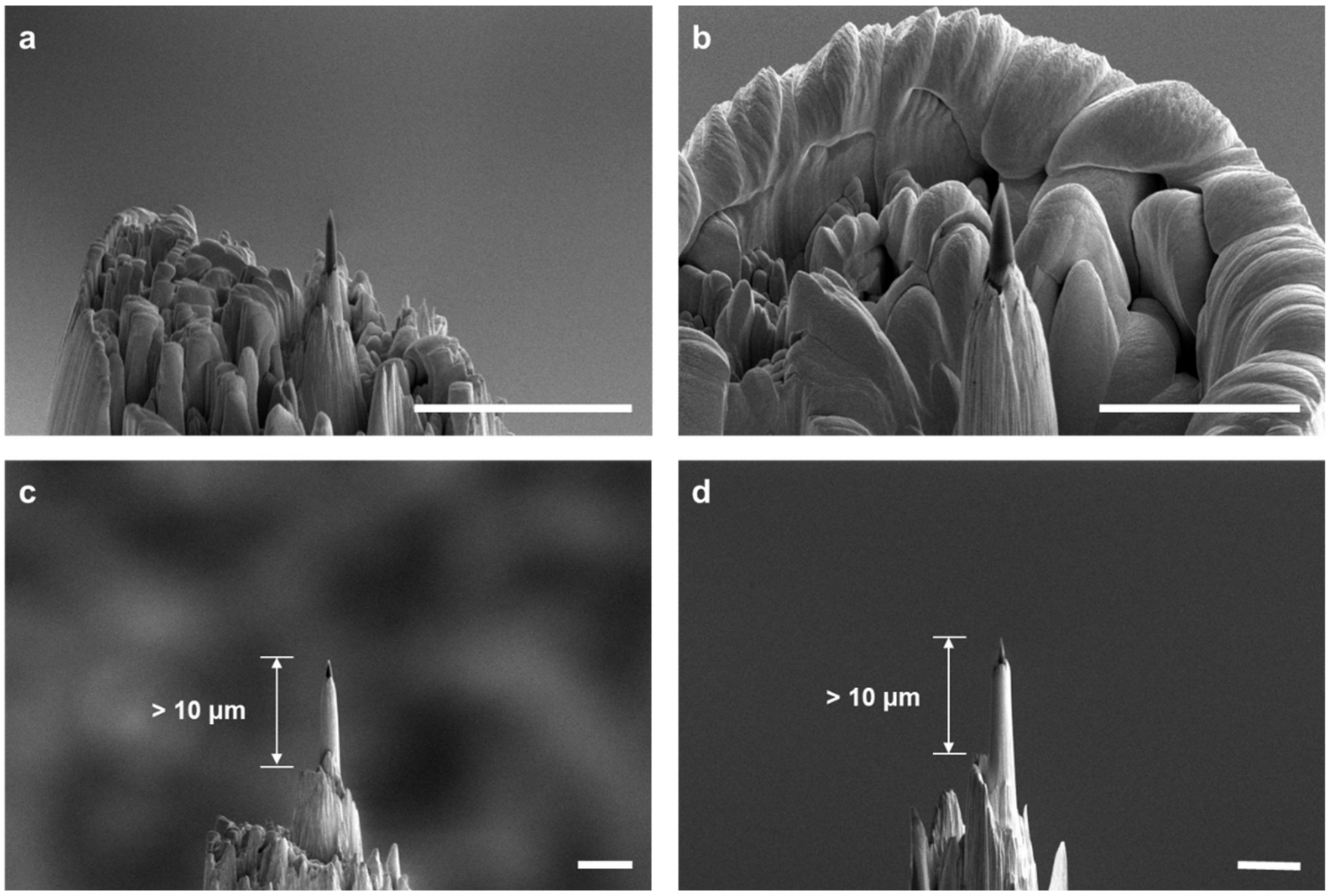

***Figure S1.1****. Requirement of an exposed single-apex geometry in the MicroCup-based workflow. (a,b) Early 5CB preparation trials showing organic apices formed close to neighboring W supports. Despite the apparently suitable geometry, these specimens did not produce stable molecular signals during APT measurements, indicating that nearby metallic protrusions can perturb the local electric field and interfere with evaporation from the organic apex. (c,d) Optimized specimens in which the frozen organic apex was positioned more than 10 µm above the nearest significant secondary protrusion. This exposed single-apex geometry minimizes competing field concentration and enables stable field evaporation from the intended organic molecular region. Scale bars: 10 µm.*

Therefore, in the optimized workflow, the frozen organic apex was deliberately positioned well above the surrounding metal support. For all successfully measured specimens, the distance between the organic apex and the nearest secondary significant protrusion was kept larger than 10 µm (**Figure S1.1**c-d). This exposed apex geometry minimizes competing field concentration and enables stable field evaporation from the intended organic molecular region.

**S2. Reference molecular structures of 8CB and 8OCB**

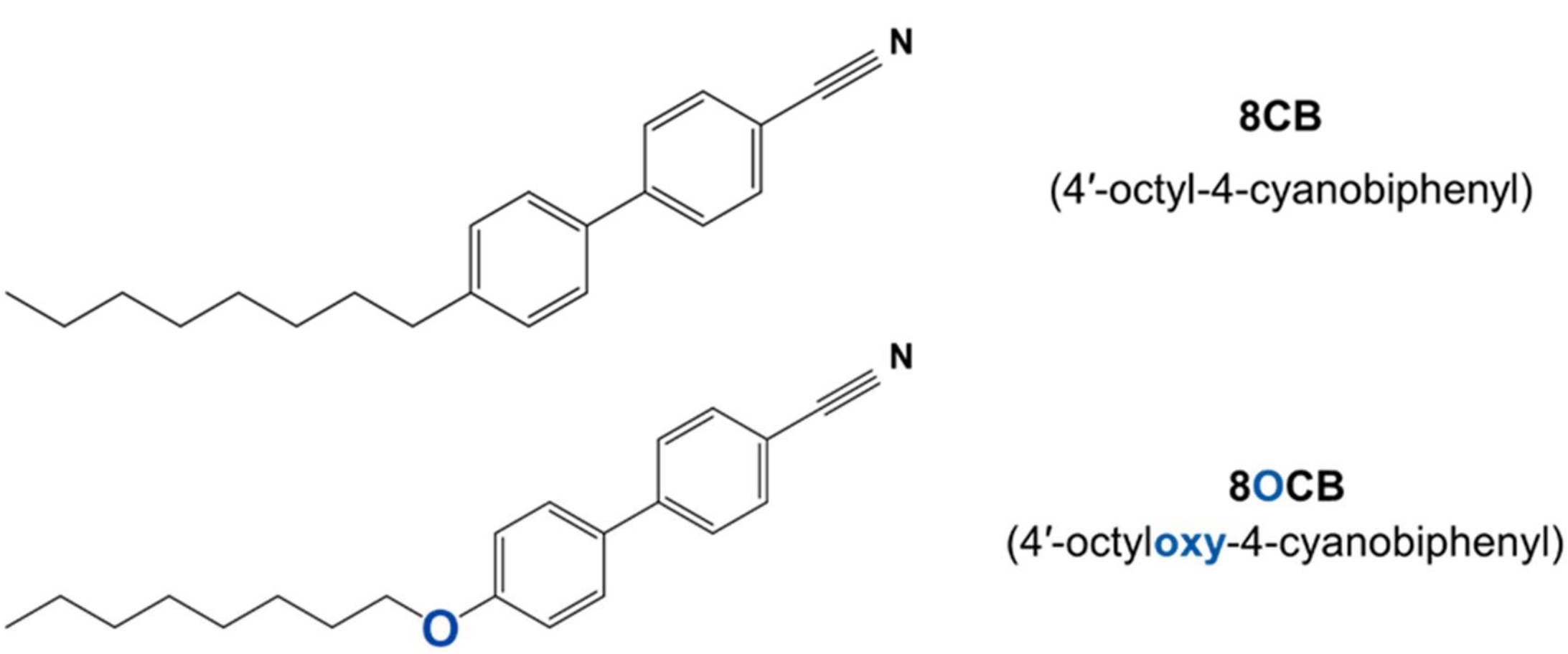

***Figure S2.1***. *Molecular structures of 8CB and 8OCB liquid crystal (LC) molecules discussed in this work. The nCB series consists of an n-carbon alkyl chain (n = 8) attached to a biphenyl core (B) terminated by a cyano group (C). In 8OCB, a single oxygen atom links the octyl chain to the cyanobiphenyl core, distinguishing it from 8CB.*

## S3. Thermal diffusion model and cooling-rate estimates

### S3.1 Cooling-rate estimation for the metallic assembly in liquid nitrogen ($LN_2$)

The cooling rate of the metallic assembly was estimated from the temperature difference and the cooling time. The temperature change was taken from room temperature (298 K) to $LN_2$ temperature (77 K), corresponding to approximately 221 K. The cooling time was measured from the moment the sample was immersed in $LN_2$ until the Leidenfrost effect disappeared (**Figure S3.1**a), which was taken as an approximate indication that the sample had reached the $LN_2$ temperature. The cooling time depended on the immersion depth. Under a low $LN_2$ level (**Figure S3.1**b), this took approximately 90 s, whereas under full immersion (**Figure S3.1**c), it took approximately 50 s. Corresponding videos have also been provided. Based on these values, the experimental cooling rate was estimated to be approximately 147-264 K/min. In the manuscript, we use the lower value as a conservative estimate.

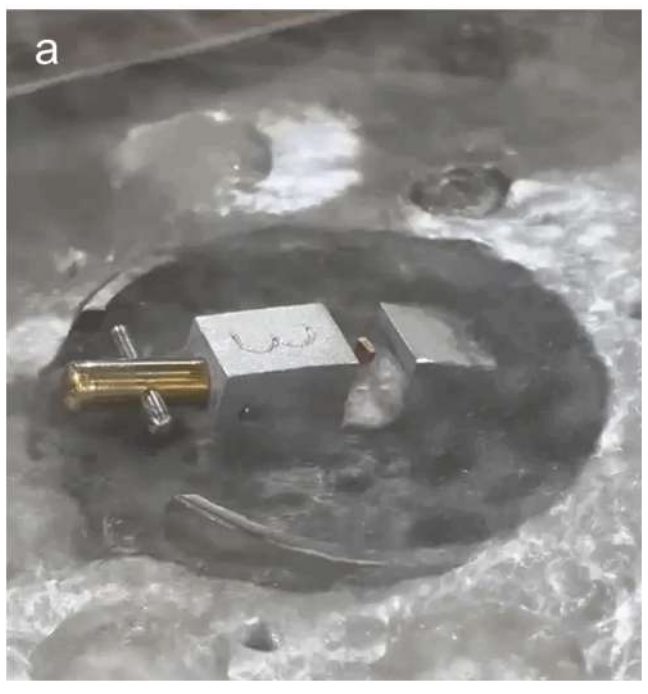

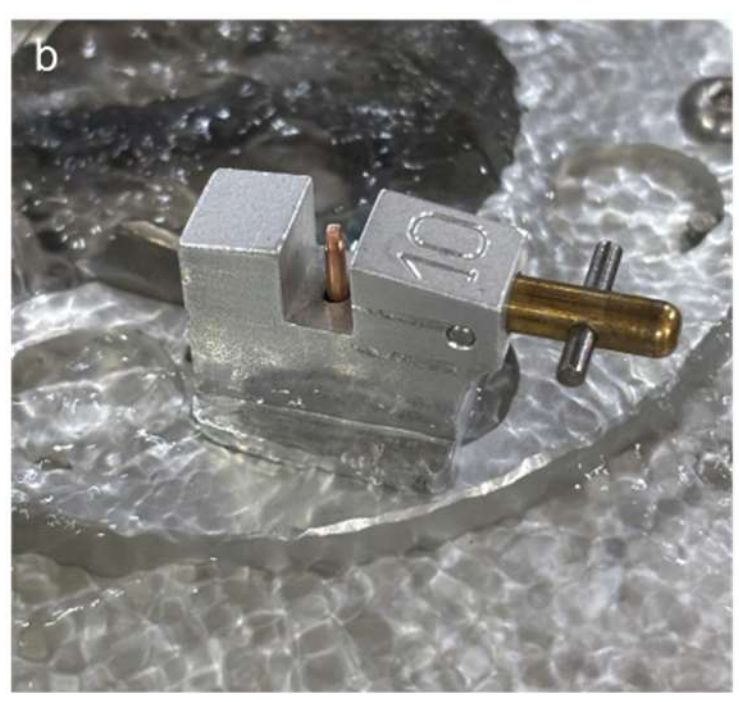

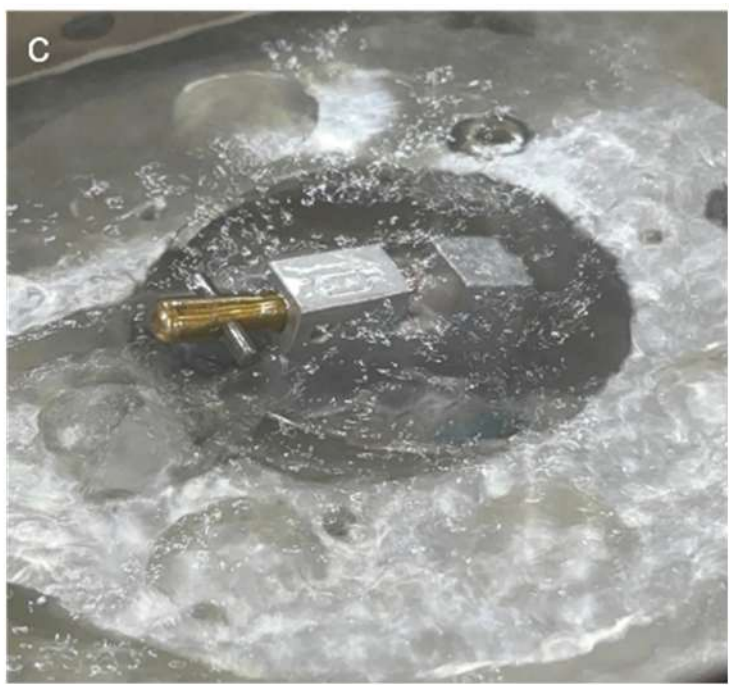

***Figure S3.1***. *(a) Illustration of the disappearance of the Leidenfrost effect. (b) Illustration of cooling under a low $LN_2$ level. (c) Illustration of cooling under full immersion in $LN_2$.*

## S3.2 Idealized estimation of the upper-bound cooling rate at the specimen scale

At the specimen scale, the W wire is immersed in liquid nitrogen and serves as the primary heat sink for the liquid specimen.

To estimate an upper bound for the achievable cooling rate, an idealized scenario is considered in which (i) the Leidenfrost effect is neglected, and (ii) heat transfer through the W wire is assumed to be sufficiently efficient that thermal resistance within the wire does not limit cooling of the liquid. Under these deliberately idealized assumptions, the early-stage cooling process can be approximated by one-dimensional thermal diffusion from the liquid nitrogen to the droplets. The characteristic diffusion depth $d$ after a time $t$ follows

$$d = \sqrt{2\alpha_l t},$$

where $\alpha_l$ represents the thermal diffusivity of investigated liquid. **Table S3.1** listed the thermal diffusivity of water, 4'-Octyl-4-cyanobiphneyl (8CB) and 4'-Octyloxy-4-cyanobiphenyl (8OCB) used in cooling rate calculation.

***Table S3.1**. Thermal Diffusivity of water, 8CB and 8OCB.*

| Name | Water(James, 1968) | 8CB(Yilmaz & Yildirim, 2009) | 8OCB* |
|---|---|---|---|
| Thermal Diffusivity ($10^{-7}$ m$^2$/s) | 1.4 | 0.77 | 1.6 |

** For 8OCB, no direct literature values of thermal diffusivity are available. The value listed here was therefore estimated from reported thermal conductivity data, combined with representative density and specific heat capacity values*(Schick et al., 2000).

Therefore, cooling from 298 K to 77 K corresponds to an effective cooling rate,

$$v_{exp} = \frac{\Delta T}{t} = \frac{2\alpha_l \Delta T}{d^2}.$$

Considering temperature difference $\Delta T$ is 221 K, the corresponding idealized cooling rates as a function of cooling depth can be estimated, which provides an upper bound for the experimentally accessible quench rates (**Figure 5**b).

**S4. Demonstration of reproducibility of the MicroCup method.**

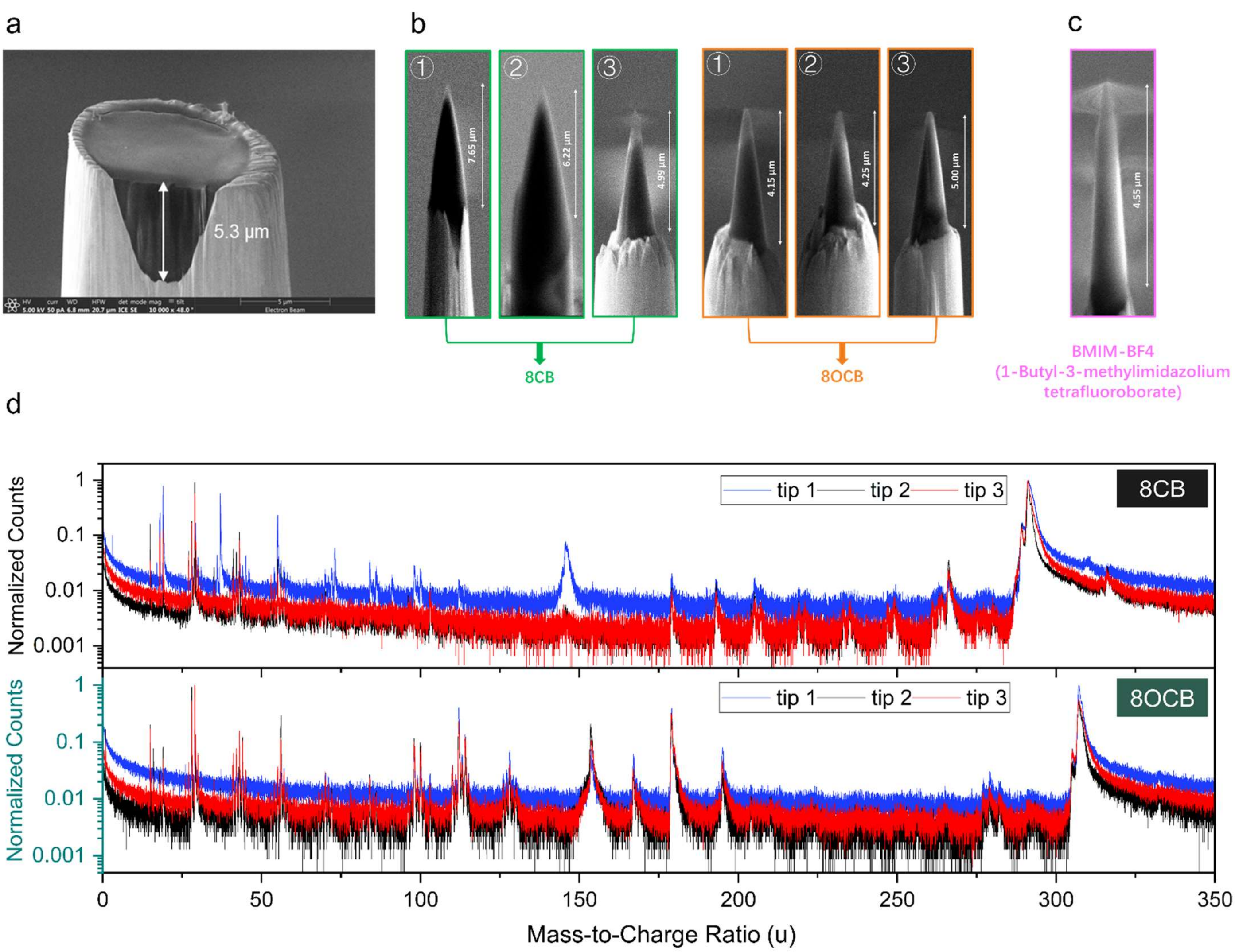

***Figure S4.1**. Reproducibility and length control enabled by the MicroCup-based preparation strategy. (a) SEM image showing direct verification of the MicroCup depth by cross-sectional exposure during specimen preparation, with an observed depth of approximately 5.3 µm. (b) SEM images of three 8CB specimens prepared using the earlier cryo-post workflow*(Meng et al., 2026) *and three 8OCB specimens prepared using the MicroCup workflow. The 8CB specimens show a broader variation in final specimen length. In contrast, the 8OCB specimens exhibit comparable final lengths of between 4 and 5 µm. This length range is close to the target geometry defined based on 8CB specimen③. (c)SEM image of an ionic-liquid specimen prepared using the MicroCup workflow. The specimen shows a similar final length of approximately 4.55 µm, demonstrating that the MicroCup geometry can provide reproducible length control for different classes of organic liquids. (c) Reproducible mass spectra obtained from pure 8CB specimens and pure 8OCB specimens in the pure smectic-like phase, indicating that the controlled specimen geometry enabled stable APT measurements and reproducible molecular signals.*

**S5. Quantification of intact molecular ion preservation and chemical stoichiometry in support information.**

To quantify intact molecular ion preservation, we first determined the event contribution of each detected species from the corresponding peak areas in the mass spectrum. These peak areas were extracted using the theoretical fitting routine implemented in APyT module(Eich, 2025), which considers signal broadening, isotopic patterns, and background subtraction. Compared with manual peak integration commonly used in conventional atom probe analysis, this approach minimizes subjective uncertainty.

The general peak profile $p(x)$ could be expressed as:

$$p(x) = \sum_j I_j \sum_i \frac{\xi_{ij}}{(x_0^{ij})^{\gamma}} a\left(x_t^{ij}\right) d\left(x_t^{ij}\right) + \frac{B}{\sqrt{x}},$$

where $j$ indexes the molecular species with intensity parameter $I_j$, and $i$ indexes the isotopes of molecule $j$. The isotope abundance is given by $\xi_{ij}$, and the denominator $x_0^{ij}$ represents the nominal mass-to-charge position of isotope $i$ in molecule $j$. The exponent $\gamma$ accounts for peak broadening in the time-of-flight mass spectrum and was fixed to 0.5, corresponding to a constant peak width in time-of-flight space. $B$ denotes the baseline parameter.

The term $x_t^{ij}$ represents the transformation into a normalized peak shape and is defined as:

$$x_t^{ij} = \frac{x - x_0^{ij}}{(x_0^{ij})^{\gamma}}.$$

The rising edge of the peak is described by the activation function

$$a(x) = \frac{1}{2}\left(\operatorname{erf}\left(\frac{\sqrt{\pi}}{2} x\right) + 1\right),$$

while the trailing edge is modeled by a simple exponential function

$$d(x) = \exp\,(-x).$$

The 5CB measurement is taken here as an example. As shown in **Figure S5.1**, the experimental mass spectrum (blue curve) is well reproduced by the fitted profile (orange curve). For molecular peak groups, the fitting procedure simultaneously considers isotopic combinations and peak broadening, enabling robust identification and quantitative integration of overlapping molecular signals. The event contribution of each determined peak is obtained, which corresponds the percentage of each determined peak area.

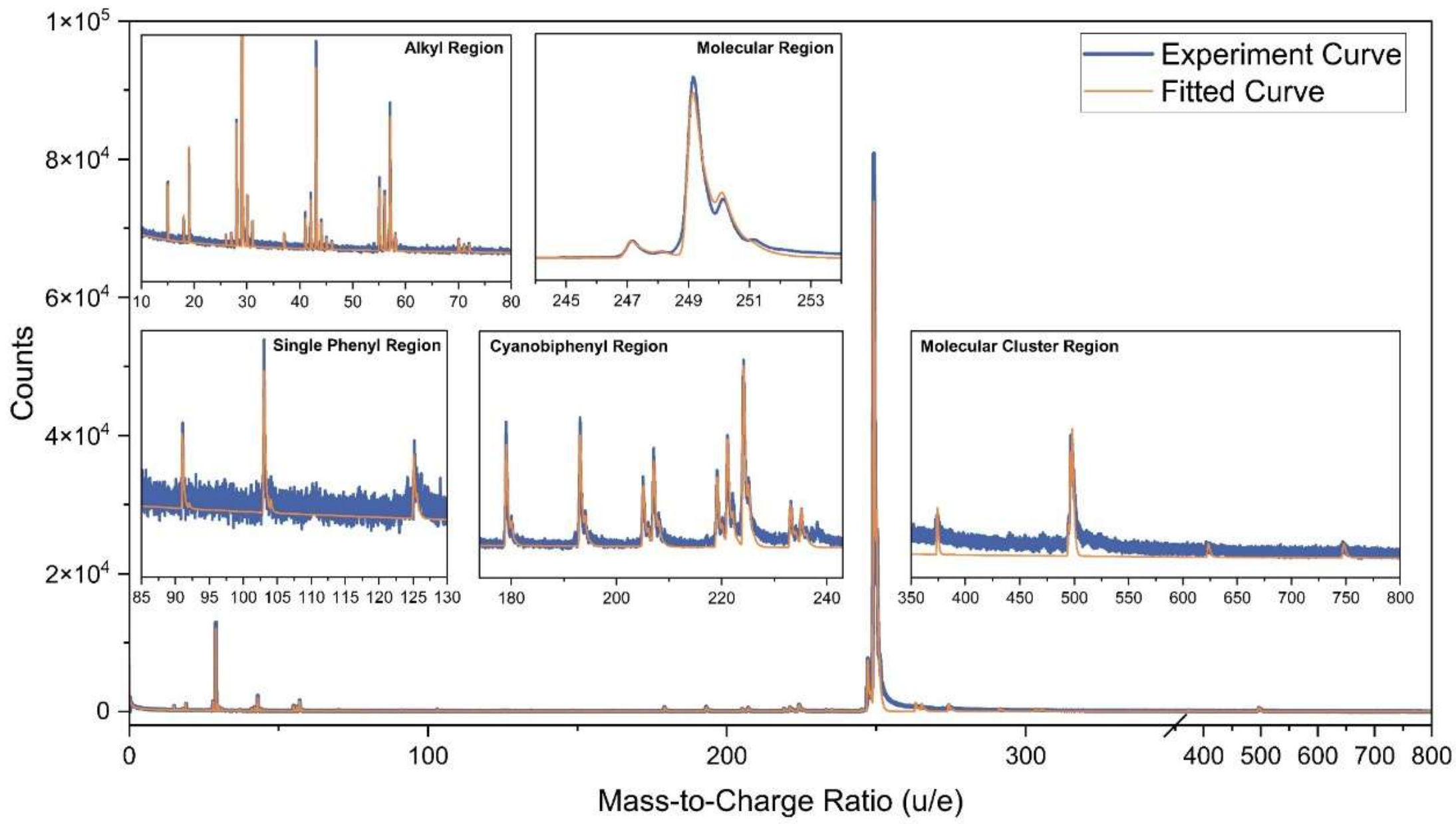

***Figure S5.1*** *Experimental mass spectrum (blue) and corresponding fitted profile (orange) obtained using the theoretical peak fitting model. The fitting reproduces the signal across the full mass-to-charge range, including the alkyl fragment, single phenyl, cyanobiphenyl, intact molecular ion, and molecular cluster regions. Insets highlight representative mass windows, demonstrating accurate peak shape reproduction, isotopic pattern matching, and background subtraction over both fragment and molecular signals.*

However, event contribution alone does not fully describe molecular preservation, because different fragment species contain different numbers of atoms. To account for this, an element-normalized analysis was performed. After determining the event contribution $A_i$ of each fitted species, the number of atoms associated with that signal was calculated as $A_i \cdot n_{X,i}$, where $n_{X,i}$ is the number of atoms of element $X$ contained in species $i$. The element-normalized molecular preservation, $P_{molecule}$, was then defined as the ratio between the total number of atoms originating from intact molecular ions and the total number of atoms from all detected species:

$$P_{molecule} = \frac{\sum_{i \in molecule} A_i \cdot n_{X,i}}{\sum_{i \in all\ species} A_i \cdot n_{X,i}}.$$

Using this approach, the intact molecular preservation was calculated for both 8CB and 8OCB in the smectic-like regions (**Table S5.1**). Based on carbon normalization, the molecular preservation is approximately 91% for 8CB and 71 % for 8OCB, consistent with the value reported in **Figure 6**f of the main text.

***Table S5.1*** *Molecular fractions (%) based on C, H, N and O atoms*

| *8CB* | | | | | | | | | | | |
|---|---|---|---|---|---|---|---|---|---|---|---|
| *Specimen 1* | | | *Specimen 2* | | | *Specimen 3* | | | *Average* | | |
| *C* | *H* | *N* | *C* | *H* | *N* | *C* | *H* | *N* | *C* | *H* | *N* |
| *93.28* | *93.76* | *94.17* | *90.75* | *91.52* | *92.37* | *90.29* | *92.10* | *91.88* | *91.44* | *92.46* | *92.81* |
| | | | | | | *Standard Deviation* | | | *1.31* | *0.95* | *0.98* |

| *8OCB* | | | | | | | | | | | | | | | |
|---|---|---|---|---|---|---|---|---|---|---|---|---|---|---|---|
| *Specimen 1* | | | | *Specimen 2* | | | | *Specimen 3* | | | | *Average* | | | |
| *C* | *H* | *N* | *O* | *C* | *H* | *N* | *O* | *C* | *H* | *N* | *O* | *C* | *H* | *N* | *O* |
| *70.88* | *70.88* | *69.31* | *82.36* | *72.63* | *72.97* | *72.36* | *83.09* | *69.83* | *71.29* | *73.08* | *74.23* | *71.11* | *71.71* | *71.58* | *79.98* |
| | | | | | | | | *Standard Deviation* | | | | *1.16* | *0.90* | *1.63* | *4.01* |

As an additional internal consistency check, the molecular preservation was independently evaluated using different elemental normalizations. For 8CB, C-, H-, and N-based normalization were compared, while for 8OCB, C-, H-, N-, and O-based normalization were considered. As shown in **Table S5.1**, normalization based on C, H, and N yields comparable preservation fractions within experimental uncertainty for both 8CB and 8OCB. In contrast, O-based normalization in 8OCB gives a higher molecular preservation of approximately 80%.

One possible explanation is that oxygen-containing fragments are more susceptible to neutral loss and therefore contribute less to the detected signal. As a result, O-based normalization gives a higher apparent molecular preservation. Although this could further support the effectiveness of our workflow in minimizing beam-induced molecular damage, we conservatively retain the lower value of approximately 70% in the main text to avoid overstatement.

**S6. Detailed interpretation of 8CB and 8OCB mass spectra.**

***Table S6.1***. *Detailed interpretation of peaks observed in the mass spectrum of 8CB.*

| Pos.$^{exp}$ (u) | Interpretation(Pos.$^{Theo}$) (u) | Assumed Origin | Other Major Possible Interpretation |
|---|---|---|---|
| **1.01** | $H^+$ (1.01) | - | - |
| **12.00** | $C^+$ (12.00) | - | - |
| **14.03** | $N^+$(14.00) | - | $CH_2^+$(14.02) |
| **15.02** | $CH_3^+$(15.02) | Methyl Fragment | $NH^+$(15.01) |
| **18.01** | $OH_2^+$(18.01) | Water | $NH_4^+$(18.03) |
| **19.01** | $OH_3^+$(19.02) | Contamination | $NH_5^+$(19.04) |
| **26.02** | $C_2H_2^+$(26.02) | Ethyl Fragments($C_2H_x^+$) | $CN^+$(26.00) |
| **27.03** | $C_2H_3^+$(27.02) | | $CNH^+$(27.01) |
| **28.03** | $C_2H_4^+$(28.03) | | $CNH_2^+$(28.02) |
| **29.04** | $C_2H_5^+$(29.04) | | - |
| **30.04** | $C_2H_6^+$(30.05) | | - |
| **31.05** | $C_2H_7^+$(31.05) | | - |
| **37.01** | $C_3H_1^+$(37.01) | Propyl Fragments($C_3H_x^+$) | $H_5O_2^+$(37.03), $C_6H_2^{++}$(37.01) |
| **39.02** | $C_3H_3^+$(39.02) | | - |
| **40.03** | $C_3H_4^+$(40.03) | | - |
| **41.04** | $C_3H_5^+$(41.04) | | - |
| **42.05** | $C_3H_6^+$(42.04) | | - |
| **43.05** | $C_3H_7^+$(43.05) | | - |
| **44.05** | $C_3H_8^+$(44.06) | | - |
| **46.08** | $C_3H_{10}^+$(46.08) | | - |
| **52.03** | $C_4H_4^+$(52.03) | Butyl Fragments($C_4H_x^+$) | - |
| **53.04** | $C_4H_5^+$(53.04) | | - |
| **54.04** | $C_4H_6^+$(54.05) | | - |
| **55.05** | $C_4H_7^+$(55.05) | | - |
| **56.06** | $C_4H_8^+$(56.06) | | - |
| **57.07** | $C_4H_9^+$(57.07) | | - |
| **58.08** | $C_4H_{10}^+$(58.08) | | |
| **67.05** | $C_5H_7^+$(67.05) | Pentyl Fragments($C_5H_x^+$) | - |
| **68.06** | $C_5H_8^+$(68.06) | | - |
| **69.07** | $C_5H_9^+$(69.07) | | *$Ga^+$(68.93, rejected) |
| **70.08** | $C_5H_{10}^+$(70.08) | | - |
| **71.09** | $C_5H_{11}^+$(71.09) | | - |
| **78.05** | $C_6H_6^+$(78.05) | Single Phenyl Group | $C_6H_6^+$ from hexyl chain |
| **81.07** | $C_6H_9^+$(81.07) | Hexyl Fragments($C_6H_x^+$) | - |
| **82.08** | $C_6H_{10}^+$(82.08) | | - |
| **83.09** | $C_6H_{11}^+$(83.08) | | - |
| **84.09** | $C_6H_{12}^+$(84.09) | | - |
| **91.04** | $C_6H_5N^+$(91.04) | $(C_6H_5+N)^+$ | $C_7H_7^+$(91.06) |
| **96.09** | $C_7H_{12}^+$(96.09) | Heptyl Fragments($C_7H_x^+$) | - |

| Pos.$^{exp}$ (u) | Interpretation (Pos.$^{Theo}$) (u) | Assumed Origin | Other Major Possible Interpretation |
|---|---|---|---|
| **97.10** | $C_7H_{13}^+$(97.10) | Heptyl Fragments($C_7H_x^+$) | $C_{21}H_{25}N^{+++}$(97.07) |
| **103.04** | $C_7H_5N^+$(103.04) | $(C_6H_5+CN)^+$ | - |
| **112.15** | $C_8H_{16}^+$(112.13) | $C_8H_x^+$ Octyl Fragment | - |
| **117.06** | $C_8H_7N^+$(117.06) | $(C_6H_5+CN+CH_2)^+$ | - |
| **131.07** | $C_9H_9N^+$(131.07) | $(C_6H_5+CN+C_2H_4)^+$ | - |
| **146.10** | $C_{21}H_{26}N^{++}$(146.10) | $(8CB+H)^{++}$ | - |
| **153.98** | $C_{11}H_8N^+$(154.07) | $(C_6H_5+CN+C_4H_3)^+$ | - |
| **168.08** | $C_{12}H_{10}N^+$(168.08) | $(C_6H_5+CN+C_5H_5)^+$ | - |
| **179.07** | $C_{13}H_9N^+$(179.07) | $(8CB-C_8H_{10})^+$ | - |
| **193.09** | $C_{14}H_{11}N^+$(193.09) | $(8CB-C_7H_8)^+$ | - |
| **205.09** | $C_{15}H_{11}N^+$(205.09) | $(8CB-C_6H_x)^+$ | $C_{16}H_{13}^+$(205.10) |
| **207.09** | $C_{15}H_{13}N^+$(207.10) | | - |
| **219.10** | $C_{16}H_{13}N^+$(219.10) | $(8CB-C_5H_x)^+$ | $C_{17}H_{15}^+$(219.12) |
| **221.03** | $C_{16}H_{15}N^+$(221.12) | | - |
| **233.12** | $C_{17}H_{15}N^+$(233.12) | $(8CB-C_4H_x)^+$ | $C_{18}H_{17}^+$(233.13) |
| **235.14** | $C_{17}H_{17}N^+$(235.14) | | - |
| **247.13** | $C_{18}H_{17}N^+$(247.13) | $(8CB-C_3H_x)^+$ | $C_{19}H_{19}^+$(247.15) |
| **249.15** | $C_{18}H_{19}N^+$(249.15) | | - |
| **261.15** | $C_{19}H_{19}N^+$(261.15) | $(8CB-C_2H_x)^+$ | $C_{20}H_{21}^+$(261.16) |
| **263.16** | $C_{19}H_{21}N^+$(263.17) | | - |
| **266.19** | $C_{20}H_{26}^+$(266.20) | $(8CB-CN+H)^+$ | $C_{19}H_{24}N^+$(266.19) |
| **275.13** | $C_{20}H_{21}N^+$(275.17) | $(8CB-CH_4)^+$ | $C_{21}H_{23}^+$(275.18) |
| **280.21** | $C_{20}H_{26}N^+$(280.21) | $(8CB-C+H)^+$ | - |
| **287.16** | $C_{21}H_{21}N^+$(287.17) | $(8CB-H_4)^+$ | - |
| **289.18** | $C_{21}H_{23}N^+$(289.18) | $(8CB-H_2)^+$ | - |
| **291.20** | $C_{21}H_{25}N^+$(291.20) | $8CB^+$ | - |
| **305.16** | $C_{22}H_{27}N^+$(305.21) | $(8CB+CH_2)^+$ | - |
| **314.19** | $C_{23}H_{24}N^+$(314.19) | $(8CB+C_2-H)^+$ | |
| **316.20** | $C_{22}H_{24}N_2^+$(316.19) | $(8CB+CN-H)^+$ | $C_{22}H_{26}N^+$(316.21) |
| **437.08** | $C_{63}H_{75}N_3^{++}$(436.80) | $(3\times8CB)^{++}$ | - |
| **582.57** | $C_{42}H_{50}N_2^+$(582.40) | $(2\times8CB)^+$ | - |

*__Pos.$^{exp}$__ refers to the experimentally determined peak maximum(Audi & Wapstra, 1993, 1995; Vocke Jr, 1999), while **Pos.$^{Theo}$** corresponds to the expected position derived from the natural-abundance isotopic distribution. Comparing these two positions allows us to differentiate genuine organic contributions from metallic species such as $Ga^+$.

For clearer visualization of the mass-spectral components, each chemical family was assigned a distinct color: molecular ions (yellow), alkyl fragments (orange), monophenyl derivatives (blue), cyanobiphenyl residues (green), and CN-H substitution pairs (purple). Variations in signal abundance within each category are indicated by graded shading (**Figure S6.1**a). In addition, **Figure S6.1**b provides a schematic overview of the characteristic field-evaporation pathways of 5CB to facilitate interpretation of the observed peak distributions.

| a Colors | Meaning |
|---|---|
| | Molecule Signals |
| | Alkyl Fragments |
| | Monophenyl Derivatives |
| | Cyanobiphenyl Residues |
| | CN-H Substitution Pairs |
| | Ether Derivatives |
| | Uncategorized |

b

N

***Figure S6.1***. *a) Color coding illustration for peak interpretation. b) Schematic of fragmentation behavior of 8CB.*

*Table S6.2. Detailed interpretation of peaks observed in the mass spectrum of 8OCB.*

| Pos.exp (u) | Interpretation(Pos.Theo) (u) | Assumed Origin | Other Major Possible Interpretation |
|---|---|---|---|
| **1.01** | $H^+$ (1.01) | - | - |
| **2.02** | $H_2^+$ (2.02) | - | - |
| **3.03** | $H_3^+$ (3.02) | - | - |
| **12.01** | $C^+$ (12.00) | - | - |
| **14.02** | $N^+$(14.00) | - | $CH_2^+$(14.02) |
| **15.03** | $CH_3^+$(15.02) | Methyl Fragment | $NH^+$(15.01) |
| **16.04** | $NH_2^+$(16.02) | - | - |
| **17.05** | $NH_3^+$(17.03) | - | - |
| **18.01** | $OH_2^+$(18.01) | Water Contamination | $NH_4^+$(18.03) |
| **19.01** | $OH_3^+$(19.02) | | $NH_5^+$(19.04) |
| **20.02** | $C_3H_4^+$(20.02) | Propyl Fragments ($C_3H_x^{++}$) | - |
| **21.02** | $C_3H_6^+$(21.02) | | - |
| **26.02** | $C_2H_2^+$(26.02) | Ethyl Fragments ($C_2H_x^+$) | $CN^+$(26.00) |
| **27.03** | $C_2H_3^+$(27.02) | | $CNH^+$(27.01) |
| **28** | $CO^+$(27.99) | Ether | $C_2H_4^+$(28.03) |
| **29.04** | $C_2H_5^+$(29.04) | Ethyl Fragments ($C_2H_x^+$) | - |
| **30.04** | $C_2H_6^+$(30.05) | | - |
| **31.05** | $C_2H_7^+$(31.05) | | - |
| **39.02** | $C_3H_3^+$(39.02) | Propyl Fragments ($C_3H_x^+$) | - |
| **40.03** | $C_3H_4^+$(40.03) | | - |
| **41.04** | $C_3H_5^+$(41.04) | | - |
| **42.05** | $C_3H_6^+$(42.04) | | - |
| **43.05** | $C_3H_7^+$(43.05) | | - |
| **44** | $CO_2^+$(43.99) | Ether Derivatives | $C_3H_8^+$(44.06) |
| **45.02** | $C_3H_9^+$(45.07) | Propyl Fragments ($C_3H_x^+$) | $CO_2H^+$(45.00) |
| **52.03** | $C_4H_4^+$(52.03) | Butyl Fragments ($C_4H_x^+$) | - |
| **53.02** | $C_4H_5^+$(53.04) | | - |
| **54.03** | $C_4H_6^+$(54.05) | | - |
| **55.04** | $C_4H_7^+$(55.05) | | - |
| **56.02** | $COC_2H_4^+$(56.03) | Ether Derivatives | $C_2O_2^+$(55.99), $C_4H_8^+$(56.06) |
| **57.03** | $C_4H_9^+$(57.07) | Butyl Fragments ($C_4H_x^+$) | $COC_2H_5^+$(57.03) |
| **58.03** | $C_4H_{10}^+$(58.08) | | $COC_2H_6^+$(58.04) |
| **66** | $C_5H_6^+$(66.05) | Pentyl Fragments ($C_5H_x^+$) | - |
| **67.01** | $C_5H_7^+$(67.05) | | |
| **68.03** | $C_5H_8^+$(68.06) | | - |
| **69.04** | $C_5H_9^+$(69.07) | | *$Ga^+$(68.93, rejected) |
| **70.06** | $C_5H_{10}^+$(70.08) | | - |
| **71.04** | $C_5H_{11}^+$(71.09) | | |
| **72.07** | $C_5H_{11}^+$(72.09) | | - |
| **78.04** | $C_6H_6^+$(78.05) | Single Phenyl Group | $C_6H_6^+$ from hexyl chain |

| Pos.$^{exp}$ (u) | Interpretation (Pos.$^{Theo}$) (u) | Assumed Origin | Other Major Possible Interpretation |
|---|---|---|---|
| **81.02** | $C_6H_9^+$(81.07) | Hexyl Fragments ($C_6H_x^+$) | - |
| **82.03** | $C_6H_{10}^+$(82.08) | | - |
| **83.04** | $C_6H_{11}^+$(83.08) | | - |
| **84.06** | $C_6H_{12}^+$(84.09) | | |
| **85.04** | $C_6H_{13}^+$(85.1) | | - |
| **91.02** | $C_6H_5N^+$(91.04) | $(C_6H_5+N)^+$ | $C_7H_7^+$(91.06) |
| **96.06** | $C_7H_{12}^+$(96.09) | Heptyl Fragments ($C_7H_x^+$) | - |
| **97.07** | $C_7H_{13}^+$(97.10) | | |
| **98.07** | $C_7H_{14}^+$(98.11) | | |
| **99.05** | $C_7H_{15}^+$(99.12) | | |
| **100.12** | $C_7H_{16}^+$(100.13) | | $C_{21}H_{25}N^{+++}$(97.07) |
| **103.01** | $C_7H_5N^+$(103.04) | $(C_6H_5+CN)^+$ | - |
| **110.07** | $C_8H_{14}^+$(110.11) | Octyl Fragments ($C_8H_x^+$) | - |
| **111.05** | $C_8H_{15}^+$(111.12) | | - |
| **112.10** | $C_8H_{16}^+$(112.13) | | - |
| **113.07** | $C_8H_{17}^+$(113.13) | | - |
| **114.14** | $C_8H_{18}^+$(114.14) | | - |
| **116.98** | $C_8H_7N^+$(117.06) | $(C_6H_5+CN+CH_2)^+$ | - |
| **126.07** | $C_8H_{14}O^+$(126.1) | Octyloxy Fragments ($C_8H_xO^+$) | - |
| **127.07** | $C_8H_{15}O^+$(127.11) | | - |
| **128.06** | $C_8H_{16}O^+$(128.12) | | - |
| **129.02** | $C_8H_{17}O^+$(129.13) | | - |
| **130.04** | $C_8H_{18}O^+$(130.14) | | - |
| **131.07** | $C_9H_9N^+$(131.07) | $(C_6H_5+CN+C_2H_4)^+$ | - |
| **154** | $C_{21}H_{26}NO^{++}$(154.10) | $(8OCB+H)^{++}$ | $C_{12}H_{10}^+$(154.08) |
| **167.04** | $C_{12}H_9N^+$(167.07) | $(C_6H_5+CN+C_5H_x)^+$ | - |
| **167.96** | $C_{12}H_{10}N^+$(168.08) | | - |
| **169.06** | $C_{12}H_{11}N^+$(169.09) | | - |
| **170.10** | $C_{12}H_{12}N^+$(170.10) | | - |
| **179.06** | $C_{13}H_9N^+$(179.07) | $(8OCB-OC_8H_{10})^+$ | - |
| **193.08** | $C_{14}H_{11}N^+$(193.09) | $(8OCB-OC_7H_8)^+$ | - |
| **195.03** | $C_{13}H_9NO^+$ (195.07) | $(8OCB-C_6H_x)^+$ | - |
| **204.04** | $C_{14}H_{10}N_2^+$(204.07) | $(C_{13}H_9N^+-CN+H)^+$ | - |
| **275.12** | $C_{19}H_{17}NO^+$(275.13) | $(8OCB-C_2H_x)^+$ | - |
| **277.13** | $C_{19}H_{19}NO^+$(277.15) | | - |
| **279.17** | $C_{19}H_{21}NO^+$(279.16) | | - |
| **282.14** | $C_{20}H_{26}O^+$(282.2) | $(8OCB-CN+H)^+$ | - |
| **289.18** | $C_{21}H_{23}N^+$(289.18) | $(8OCB-O-H_2)^+$ | - |
| **291.15** | $C_{21}H_{25}N^+$(291.20) | $(8OCB-O)^+$ | - |
| **295.2** | $C_{20}H_{25}NO^+$(295.19) | $(8OCB-C)^+$ | - |
| **303.16** | $C_{21}H_{21}NO^+$(303.16) | $(8OCB-H_4)^+$ | - |

| Pos.$^{exp}$ (u) | Interpretation (Pos.$^{Theo}$) (u) | Assumed Origin | Other Major Possible Interpretation |
|---|---|---|---|
| **305.18** | $C_{21}H_{23}NO^+$(305.18) | $(8OCB-H_2)^+$ | - |
| **307.19** | $C_{21}H_{25}NO^+$(307.19) | $8OCB^+$ | - |
| **332.18** | $C_{22}H_{24}N_2O^+$(332.18) | $(8OCB+CN-H)^+$ | - |

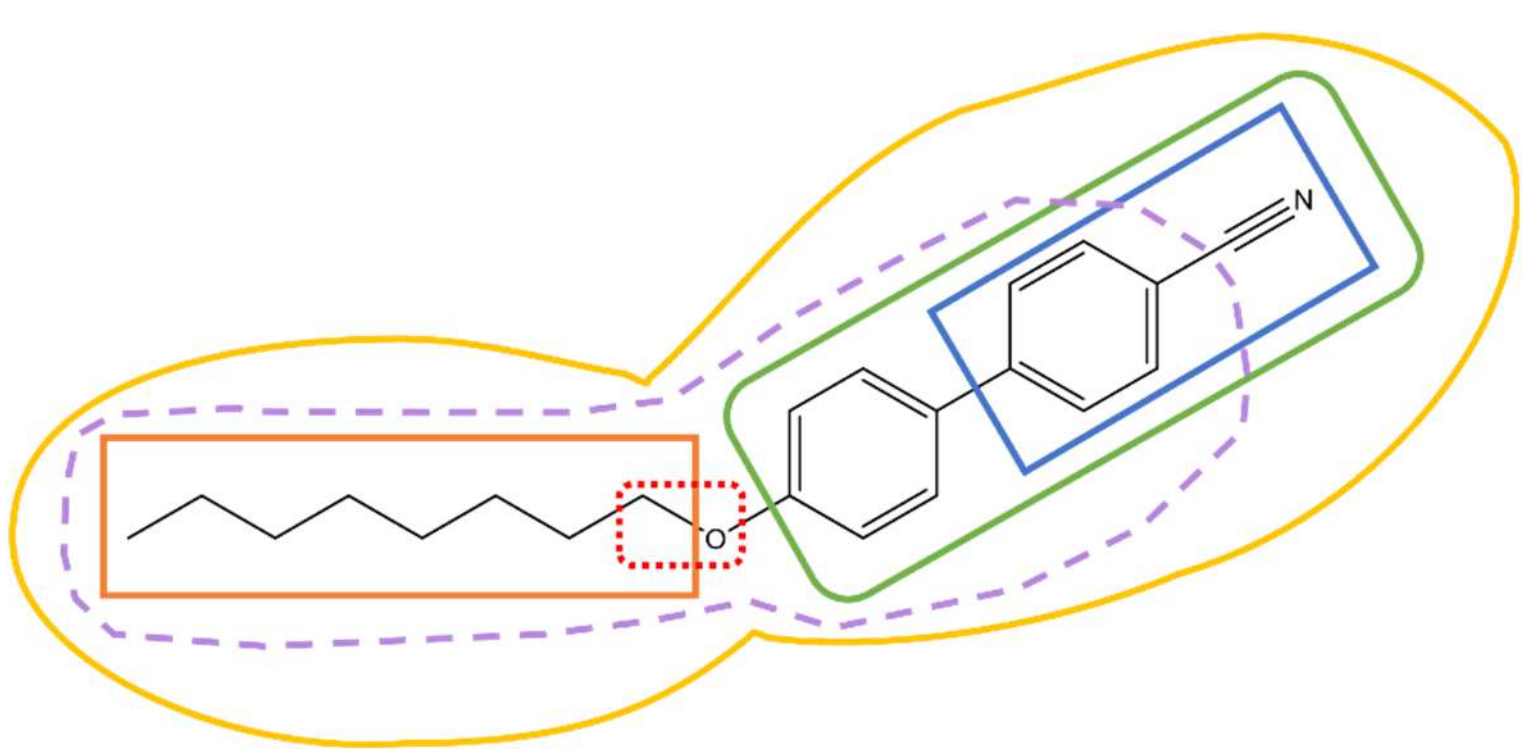

***Figure S6.2**. Schematic of fragmentation behavior of 8OCB.*

**S7. Determination of the disputed peak signals near 28, 44 and 56 u in 8OCB mass spectrum**

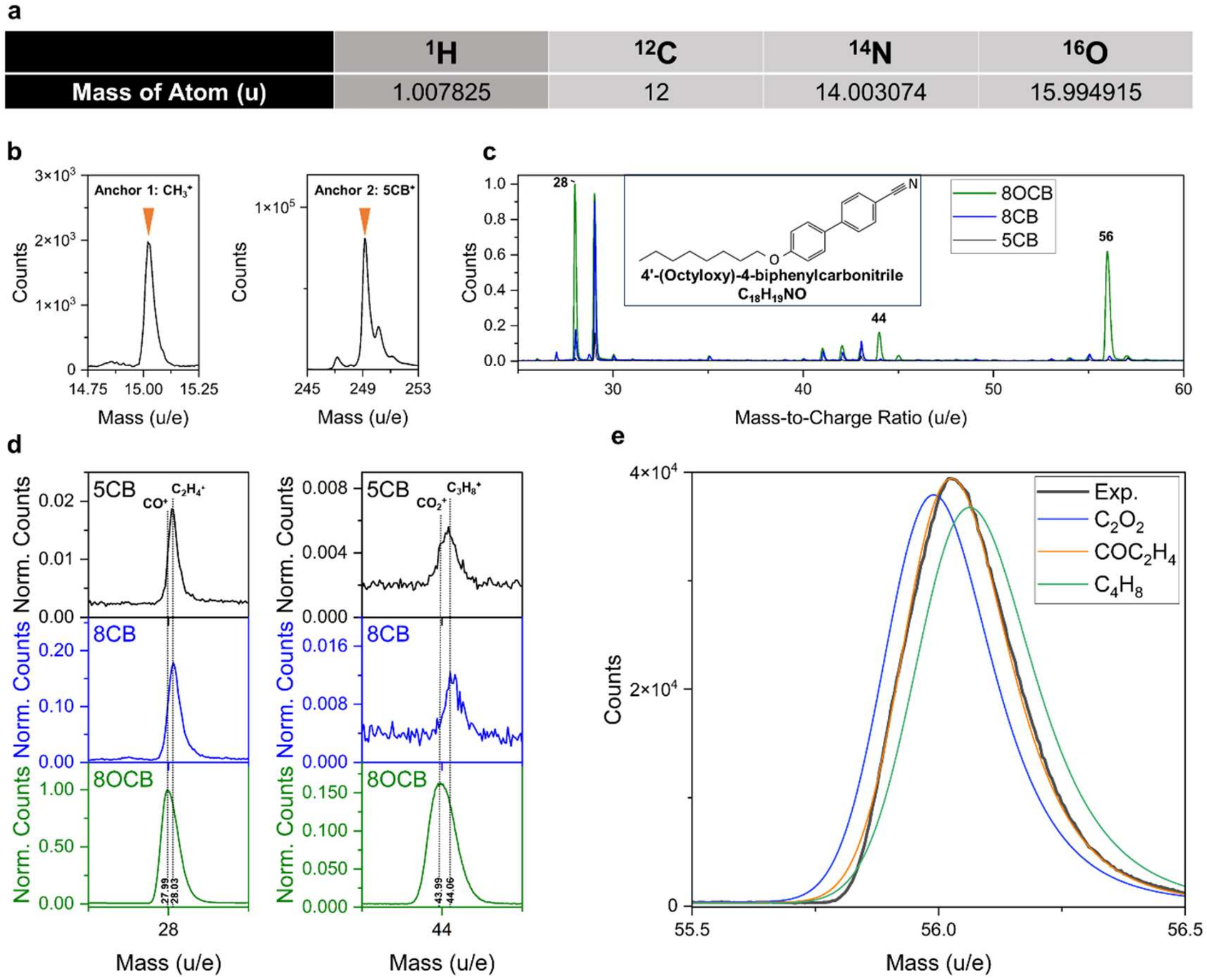

***Figure S7.1**. Resolving the disputed peaks in the 8OCB mass spectrum using mass-deficit analysis and spectral fitting. (a) Mass deficits of the most abundant isotopes in light elements* (Audi & Wapstra, 1993, 1995; Vocke Jr, 1999). *(b) Peak alignment based on two undisputed reference peaks to avoid arbitrary assignment. (c) Normalized comparison of 5CB, 8CB, and 8OCB spectra, highlighting three unresolved peaks near 28, 44, and 56 u. (d) Mass-deficit analysis confirming the assignments of the 28 and 44 u peaks.(e) Spectral fitting using the APyT mass-spectrum fitting module* (Eich, 2025) *identifying the 56 u peak.*

Our assignment of the peak near 28 u in 8OCB as being predominantly $CO^+$ is based on multiple observations.

First, the comparison among 5CB, 8CB, and 8OCB provides a clear internal reference. All three molecules contain one cyano group, whereas only 8OCB contains an additional ether oxygen linking the alkyl chain to the cyanobiphenyl core. As shown in **Figure S7.1**c, the peaks near 28, 44, and 56 u are strongly enhanced in 8OCB, while they remain much weaker in 5CB and 8CB. More specifically, the event contribution of the peak near 28 u increases from approximately 0.14% in 5CB, 1.07% in 8CB (Meng et al., 2026) to approximately 18% in 8OCB. This systematic enhancement is therefore more consistent with oxygen-related fragmentation than with fragments derived primarily from the common cyano group.

Second, our interpretation is further supported by the peak centroid positions. Before comparing 5CB, 8CB and 8OCB, the spectra were calibrated using $CH_3^+$ and the molecular peaks, without introducing any arbitrary peak shift. An example is shown in **Figure S7.1**b for the first isotopologue peak of 5CB, $^{12}C_{18}{}^{1}H_{19}{}^{14}N^+$, at 249.15 u. For the peak near 28 u, the centroid lies at approximately 28.03 u in 5CB and 8CB, but shifts to approximately 28.00 u in 8OCB. The 8OCB centroid is the closest to the theoretical mass of $CO^+$ (27.99 u), rather than those of $CNH_2^+$ (28.02 u) and $C_2H_4^+$ (28.03 u). By contrast, neighboring peaks near 27 and 29 u in all spectra remains to be around 27.03 u and 29.05 u respectively, which are consistent with $C_2H_x^+$ series. **Figure S7.1**d shows a clear shift of the peaks near 28 and 44 u among the three systems. It should be noted such small shift has also been observed in previous work and utilized for indication different peak species(Perea et al., 2016; Schmidt et al., 2018).

Moreover, **Figure S7.1**e demonstrates our spectral fitting using the APyT mass-spectrum fitting module (Eich, 2025). The experimental peaks can be optimally fitted by this theoretical curve corresponding to the $COC_2H_4^+$ species (orange line) rather than $C_2O_2^+$ species (blue line) and $C_4H_8^+$ species (green line). This further confirms the proposed assignment.

Taken together, the peaks near 28, 44, and 56 u can be consistently interpreted as oxygen-containing fragments. They are most likely derived from the ether linkage in 8OCB, with different numbers of methylene units from the adjacent alkyl chain.

## S8. Assessment of Ga contamination

First, the overall event contribution of the 69 u peak is very low, below 0.2% in both datasets (**Table S8.1**). Specifically, its contribution is 0.08% in 8CB and 0.13% in 8OCB. For comparison, the adjacent 70 u peak shows consistently higher contributions, namely 0.19% in 8CB and 0.23% in 8OCB. While the 70 u signal has a clear assignment to organic species, its higher event contribution than 69 u proves that MicroCup method effectively minimized Ga contamination.

***Table S8.1*** *Event contribution of peak signal near 69 u in 8CB and 8CB datasets*

| 8CB | | 8OCB | |
|---|---|---|---|
| 69 u | 70 u | 69 u | 70 u |
| 0.08 % | 0.19 % | 0.13 % | 0.23 % |

Second, we analyzed the spatial distribution of the 69 u signal, as now shown in **Figure S8.1**. In the 8CB specimen, the 69 u ions are distributed approximately uniformly throughout the dataset (**Figure S8.1**a). In the 8OCB specimen, the 69 u signal is in fact slightly less dense in the region evaporated at the beginning of the measurement (yellow arrow in **Figure S8.1**b) than in the later-evaporated region. This trend is opposite to what would be expected if the signal originated from implanted Ga introduced during specimen preparation.

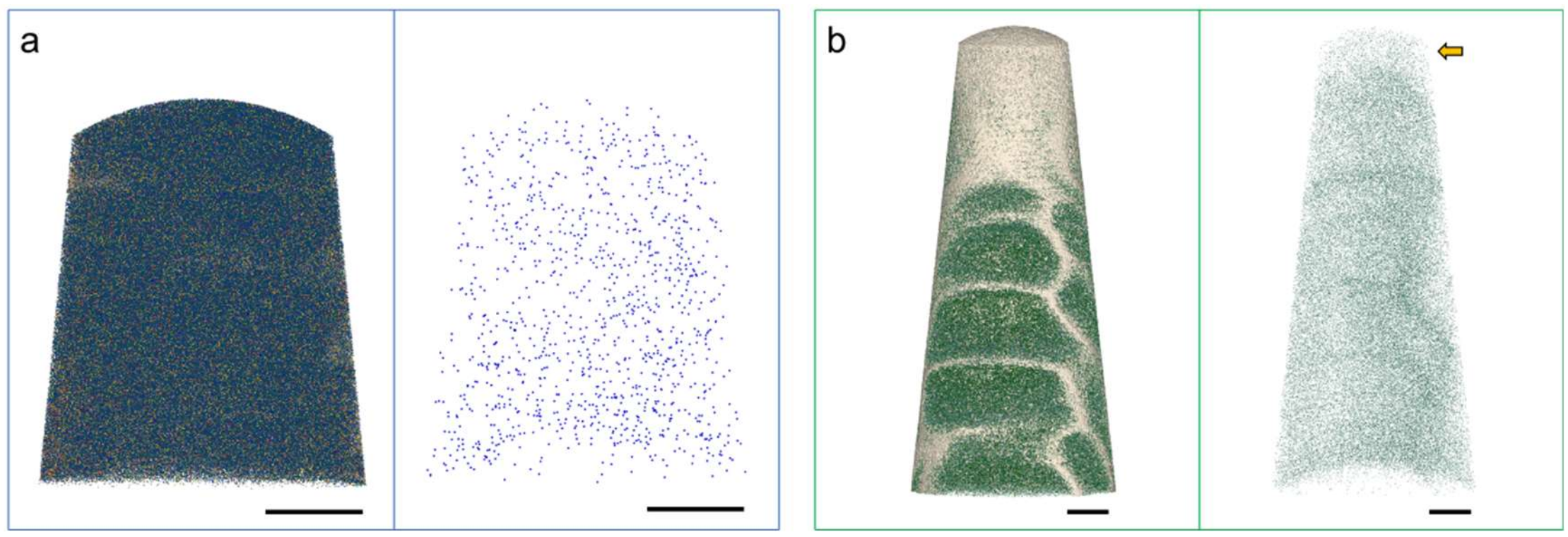

***Figure. S8.1*** *Spatial distribution of the 69 u signal in the analyzed datasets. (a) Reconstructed 8CB dataset (left) and the corresponding spatial distribution of the 69 u ions (right). (b) Reconstructed 8OCB dataset (left) and the corresponding spatial distribution of the 69 u ions (right). The yellow arrow indicates the region evaporated at the beginning of the measurement. Scale bars: 50 nm.*

Taken together, the 69 u peak should most likely originate from an organic fragment instead of Ga contamination.

**S9. Example of n-Decane analysis illustrating that a high-quality mass spectrum does not necessarily guarantee a reliable spatial reconstruction**

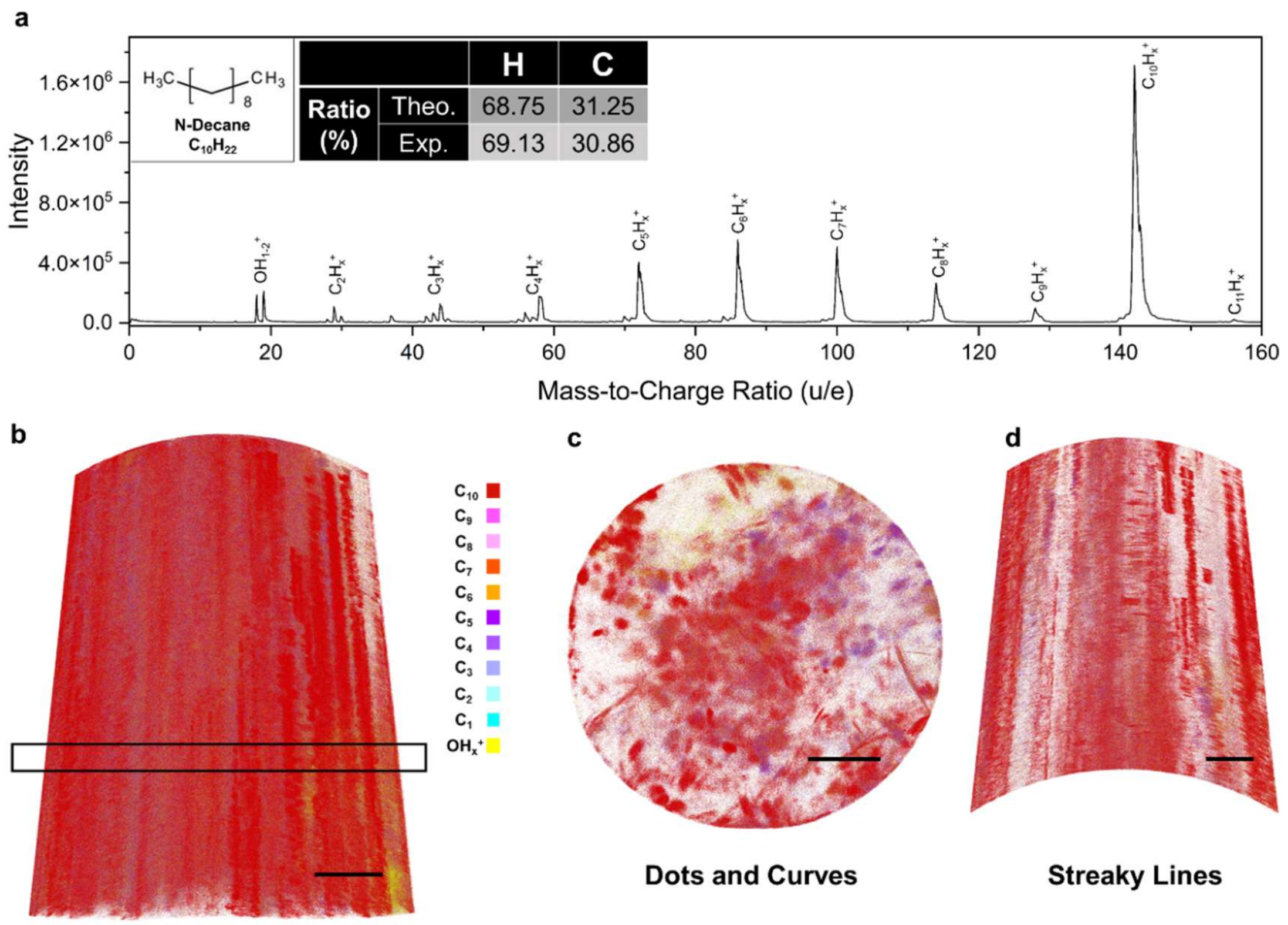

***Figure S9.1***. *Example of n-Decane analysis illustrating that a high-quality mass spectrum does not necessarily guarantee a reliable spatial reconstruction. (a)The mass spectrum shows well-defined peaks and chemical stoichiometry matching the theoretical composition. (b)The corresponding 3D reconstruction, from which a 20-nm-thick slice was extracted for detector-view analysis in (c). (c)The presence of dispersed dots and curved ion tracks, and (d) streak-like features, represent typical reconstruction artifacts in organic liquids measurements. These artefacts compromise spatial accuracy, which should be distinguished from the intrinsic structure from soft materials. Similar concerns were raised by Stender et al.* (Stender et al., 2022) *who demonstrated that even when an unshaped frozen water droplet was irradiated with high laser power, protonated water clusters could still be obtained in the mass spectrum despite severe spatial distortions. Scale bar: 50 nm.*

**S10. Supporting evidence for assigning the periodic molecular feature to a crystalline 8CB domain**

In the main text, the field evaporation behavior of different 8CB regions is discussed based on the assignment of a localized periodic molecular feature to a crystalline 8CB domain. Because this assignment is important for the subsequent interpretation, we provide here a more detailed evaluation of the origin of this periodicity.

First, we clarify that the feature is not interpreted as conventional atomic planes. In the 8CB dataset, the dominant detected species are intact molecular ions, $C_{21}H_{25}N^+$, together with related molecular fragments. Therefore, the relevant question is whether the observed periodicity corresponds to molecular crystallographic ordering in a crystalline 8CB domain.

## S10.1 Relationship between the periodic feature and local specimen curvature

In conventional APT crystallographic-plane imaging, planes are generally observed locally below crystallographic poles and are tangent to the local specimen curvature(Gault et al., 2010). As shown in **Figure S10.1**, the orientation of the periodic features aligns closely with the reconstructed surface slope. This slope acts as an approximation of the local curvature tangent, showing that the observed planes are generally parallel to the tangent of the local curvature.

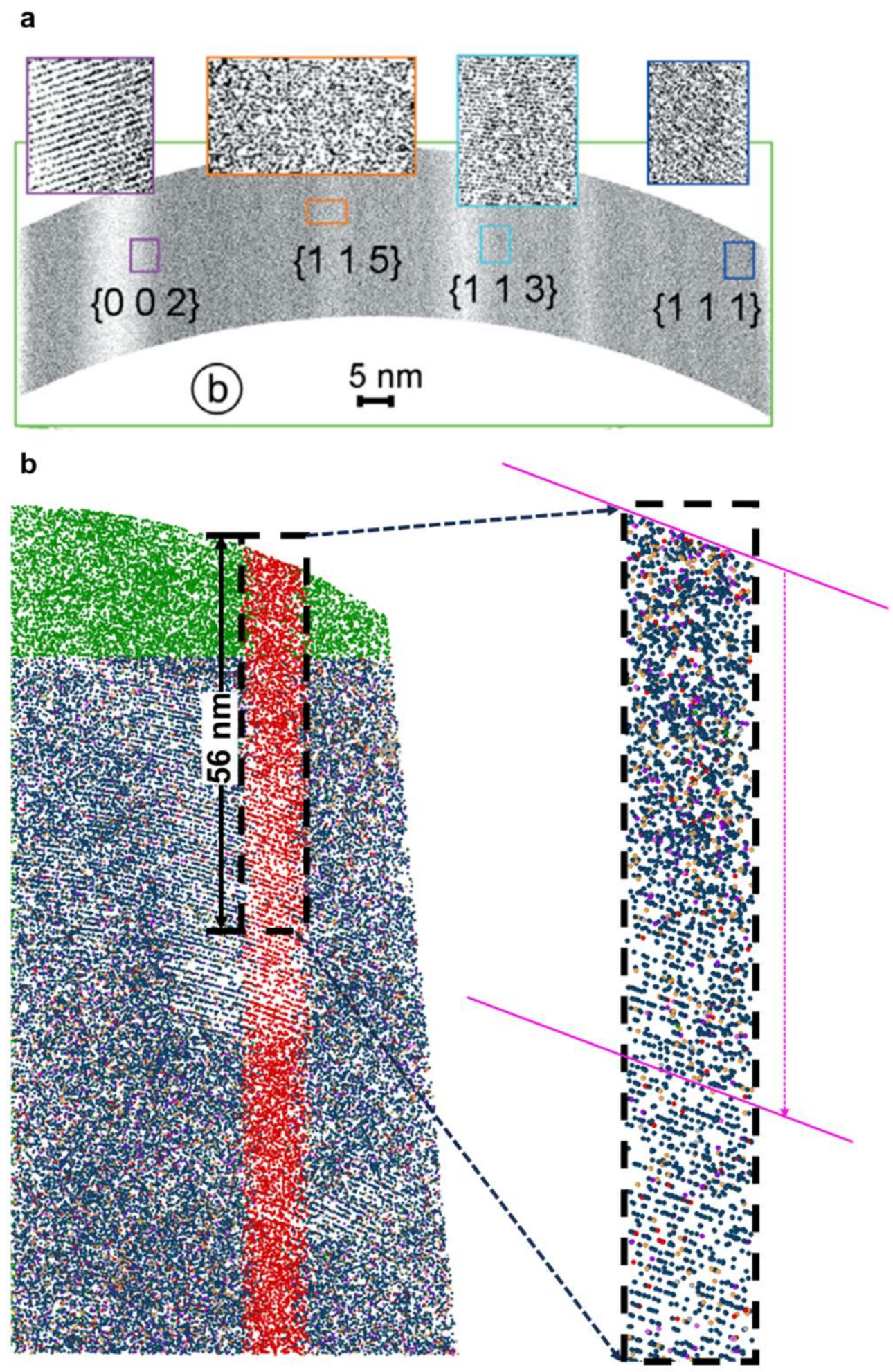

***Figure S10.1**. Revised 8CB reconstruction including the upper reconstructed region, which allows the local specimen curvature to be identified. The magnified region shows that the apparent periodic*

*molecular features are approximately parallel to the tangent of the local curvature. The length of the extracted region is approximately 56 nm.*

Second, while this is a useful approximation, the reconstructed surface slope is not the same as the local curvature tangent. To illustrate this distinction, we examined two periodic domains at the same polar angular position. These domains would originate from the same area of the reconstructed surface, sharing the same local slope. Yet, we found that their plane orientations differ slightly (**Figure S10.2**). This reinforces that the periodic features align more fundamentally with the local curvature tangent rather than just the reconstructed surface slope.

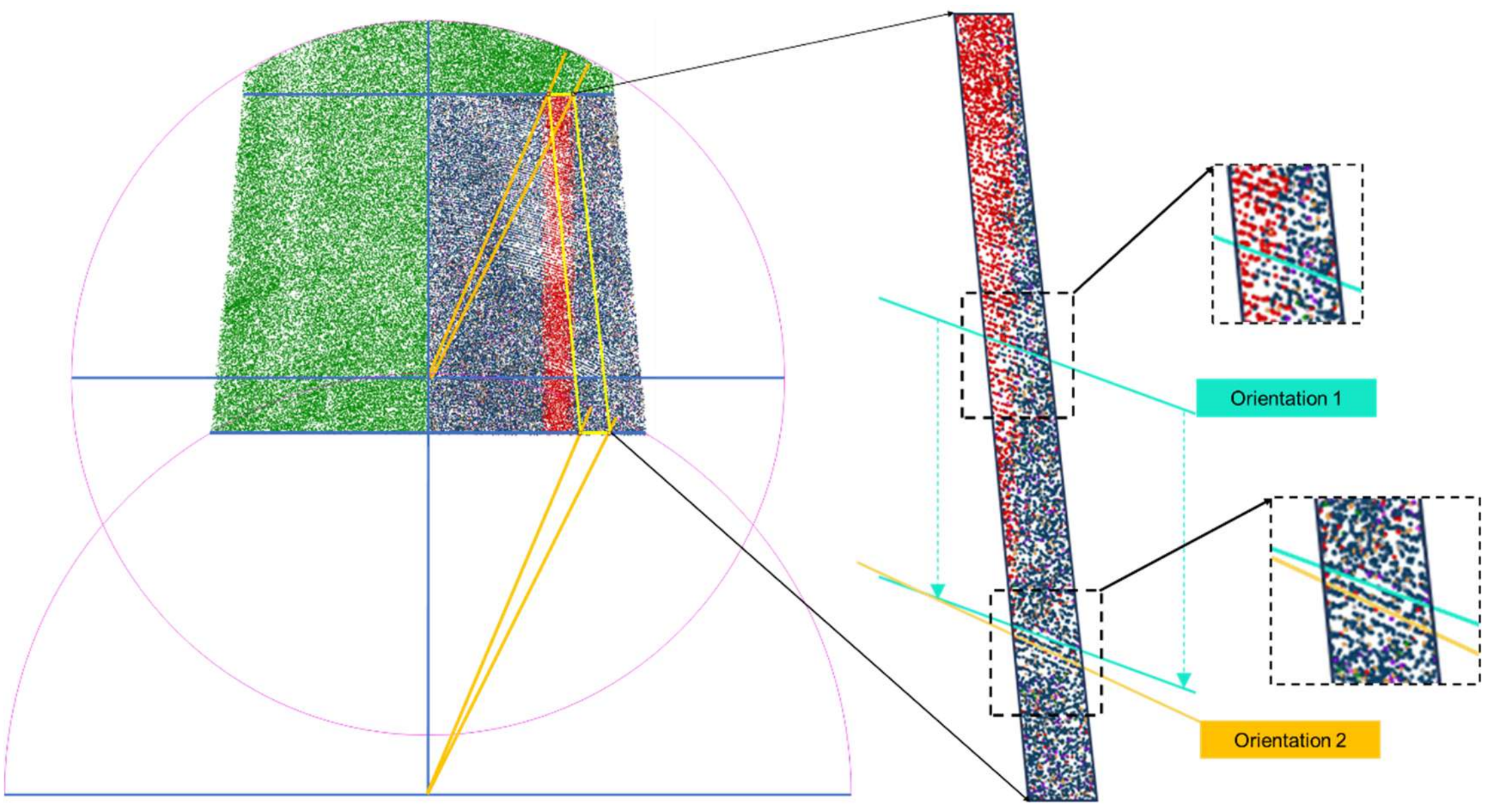

***Figure S10.2*** *Re-examination of the apparent periodic feature within region from the same polar angle position. The local orientation 1 (cyan) of the planes is slightly different from the orientation 2 (yellow) of the planes from another grain.*

### S10.2 Detector hit-map analysis during the evaporation range of the periodic region

We further inspected the detector hit maps during the evaporation range associated with the periodic molecular feature. As shown in **Figure S10.3**, over 12,200 consecutive events, from event 289,000 to 301,200, the detector view shows a clear pole-like indication with a stable ring-like ion distribution around a localized region away from the tip axis. This indicates that the periodic features were not extracted from a random part of the reconstruction, but from a limited region associated with a pole-like evaporation pattern.

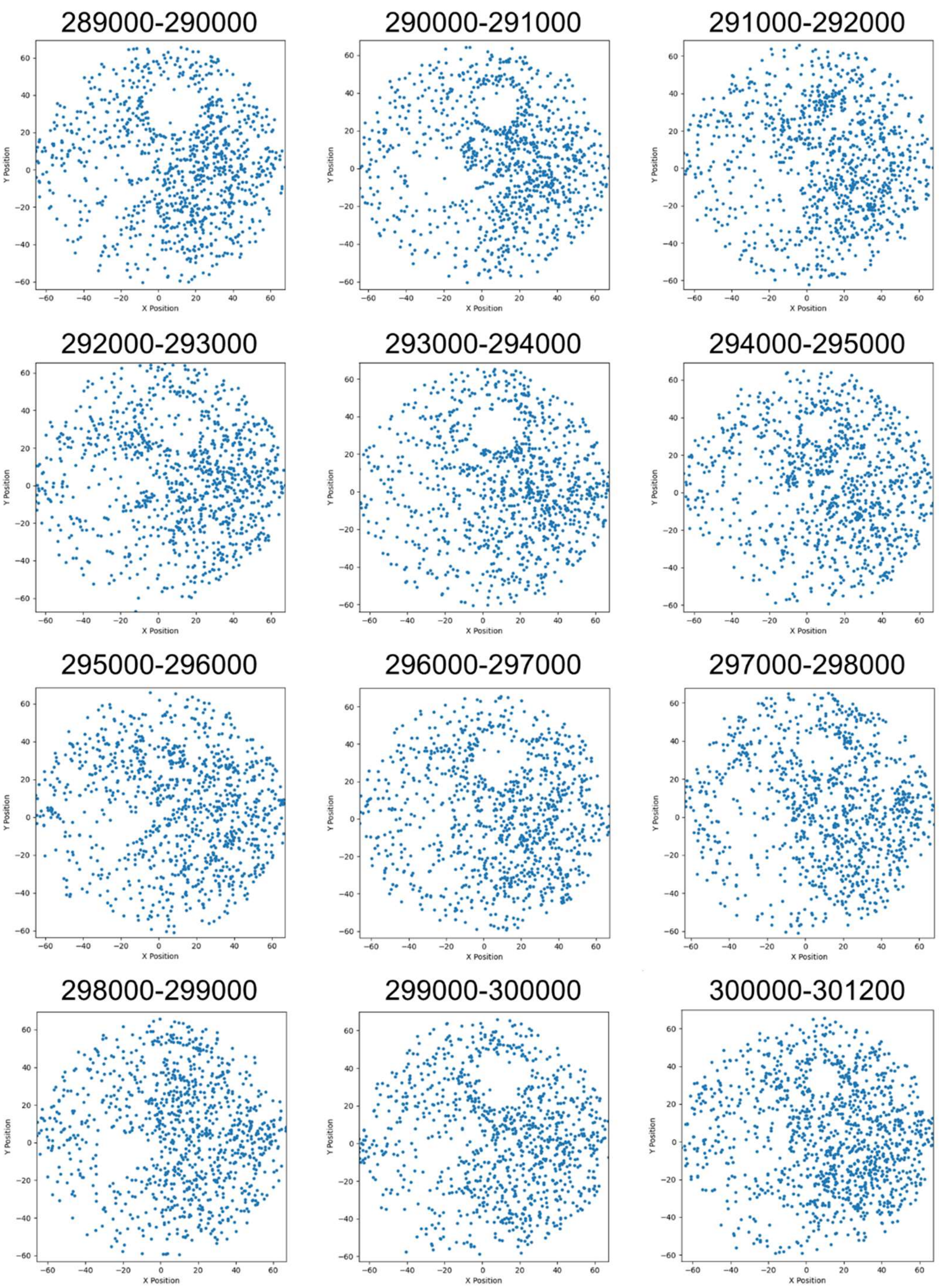

***Figure S10.3***. *Detector hit maps during the evaporation range associated with the periodic pole-like indication.*

### S10.3 z-SDM and FFT analysis of the periodic molecular feature

To evaluate whether the observed periodicity could arise purely from reconstruction or field-evaporation artefacts, we performed spatial distribution map analysis in the same grain (Meng et al., 2026). As shown in **Figure S10.4**, the z-SDM profile shows a clear one-dimensional periodic response with an average spacing of 0.731 ± 0.048 nm. The Fourier analysis further shows that this periodic response is mainly associated with molecular signals and related molecular fragments. In contrast, hydroxide extracted from the same spatial region do not show the same periodic response.

If this periodicity were a reconstruction artefact, it would lack this chemical specificity and affect all species. We are unaware of any reconstruction artefact mechanism that would produce a stable 1D spacing solely for molecular signals while leaving other signals unaffected. Therefore, a purely artefactual origin appears unlikely.

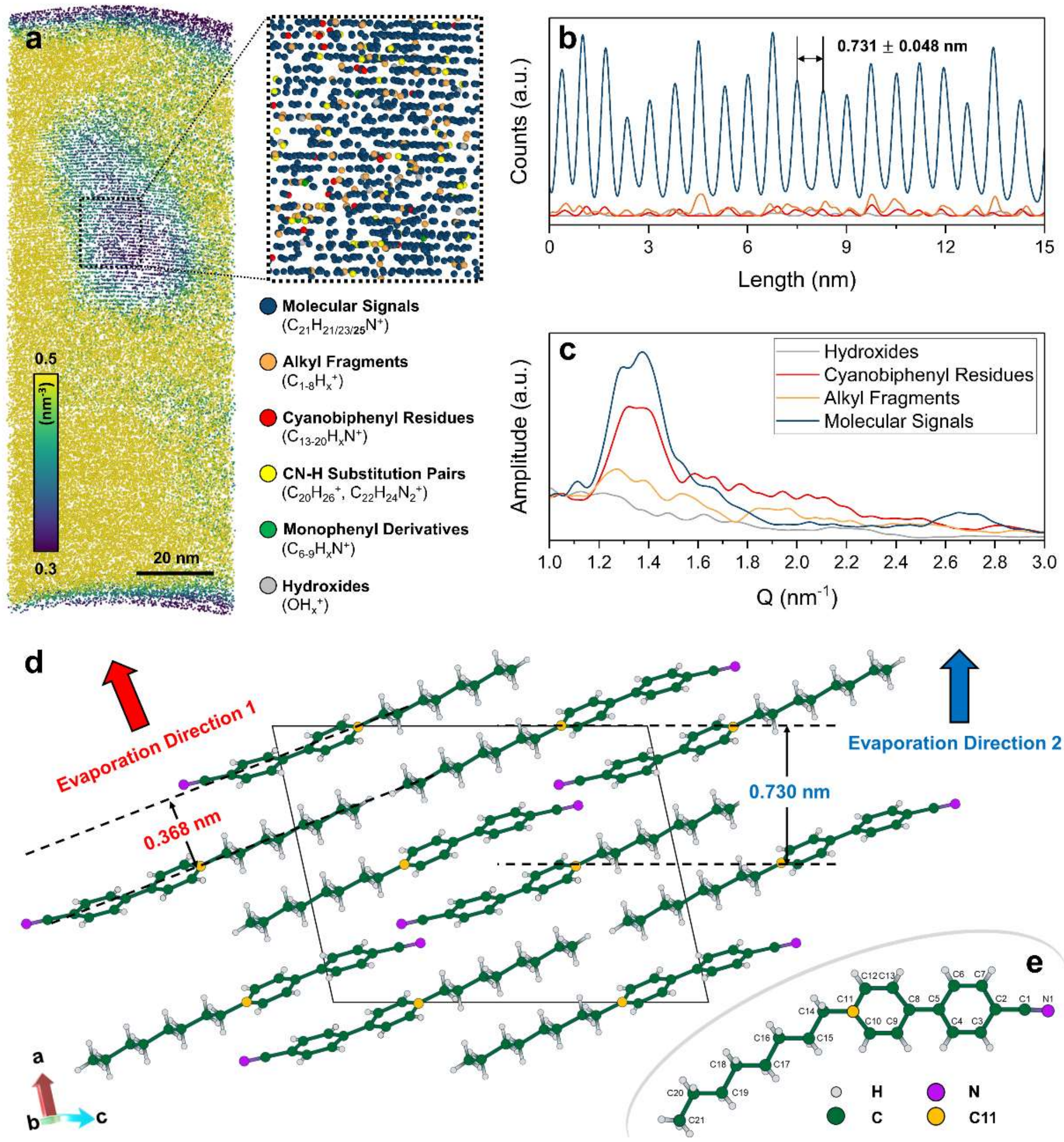

*Figure S10.4. SDM and FFT analysis of the periodic molecular feature in the 8CB dataset. (a) Local density map of a 10 nm-thick slice from the reconstructed 8CB dataset, highlighting a layered molecular feature in the low-density region. A zoomed-in view of a 17 nm-thick region along the z direction is shown on the right, with the major molecular and fragment signals color-coded. (b) z-SDM profile showing a one-dimensional periodicity with an average spacing of 0.731 ± 0.048 nm. (c) FFT analysis showing that the periodic response is mainly associated with molecular signals and related molecular fragments, whereas hydroxide/background signals do not show the same periodic correlation. (d) Schematic representation of the proposed field-evaporation direction relative to molecular orientation in the crystalline 8CB domain. (e) Molecular structure of 8CB, with the central carbon atom C11 used as a consistent molecular reference point. Reproduced from the authors' previous work*(Meng et al., 2026).

**S10.4 Distinction between smectic layering and crystalline molecular periodicity**

The interpretation of a one-dimensional periodicity in a thermotropic LC system requires particular caution, because such periodicity could in principle originate either from smectic layering or from crystalline ordering. Smectic phase is a mesophase that has molecular orientational order and one-dimensional translational order perpendicular to the layers, whereas crystalline phase has three-dimensional translational order together with molecular orientational order (**Figure S10.5**).

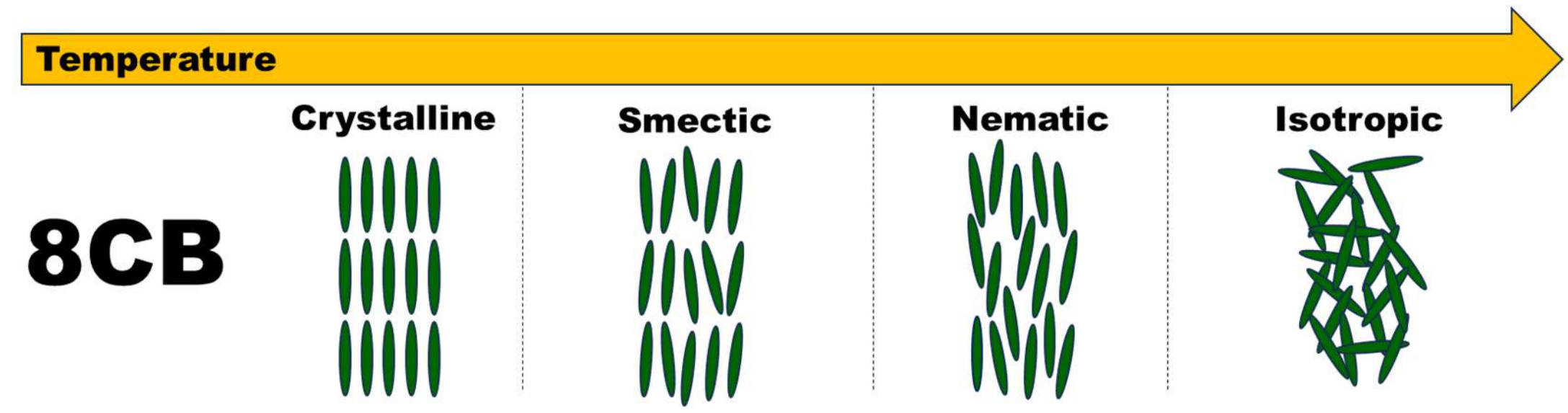

***Figure S10.5***. *Schematic distinction between crystalline, smectic, nematic, and isotropic ordering in thermotropic 8CB*.

The assignment primarily relies on the measured interplanar spacing. The measured spacing of 0.731 ± 0.048 nm is far below the reported smectic layer spacing of 8CB, 3.173 nm (Safinya et al., 1993), and cannot be reconciled with smectic layering within the constrained reconstruction parameters. This strongly suggests that the observed periodicity does not originate from the smectic layer spacing, but is more consistent with crystalline molecular ordering.

**S10.5 Comparison with the reported crystalline structure of 8CB**

To date, only one crystalline structure of 8CB has been reported (Kuribayashi & Hori, 1998). We therefore compared the experimentally measured periodic spacing with the interplanar distances expected from this crystalline structure.

Because 8CB is an anisotropic rod-like molecule, the definition of a molecular plane requires a consistent molecular reference point. Here, we used the central carbon atom C11 as a representative molecular reference point (**Figure S10.4**e). Based on this representation, the expected spacing of the crystalline (200) planes is approximately 0.730 nm (see evaporation 2 in **Figure S10.4**d), which closely matches the experimentally measured spacing of 0.731 ± 0.048 nm. This agreement supports the assignment of the observed periodicity to a crystalline 8CB molecular domain, most likely associated with the (200) periodicity. The alternative $(40\bar{2})$ direction was also considered (see evaporation 1 in **Figure S10.4**d). Although this orientation can produce a clear planar arrangement when projected onto the ac-plane, it is less likely because the corresponding evaporation direction is nearly perpendicular to the molecular long axis and would not align with the expected easy molecular dipole direction (Meng et al., 2026). It is also a higher-index direction, whereas APT crystallographic contrast is more commonly associated with low-index poles.

**S10.6 xy-SDM comparison with the proposed crystalline orientation**

We also compared the experimental xy-SDM with the projected molecular arrangement expected for the first two molecular layers of the proposed (200) crystalline orientation. Although the ion counts are too limited to claim a full two-dimensional crystallographic match, the xy-SDM shows non-random spatial correlations consistent that indicate long-range periodicity which could not be understood by an instability in the evaporation process. Still, larger single-crystalline domains would be required in future experiments for a more complete two-dimensional SDM validation..

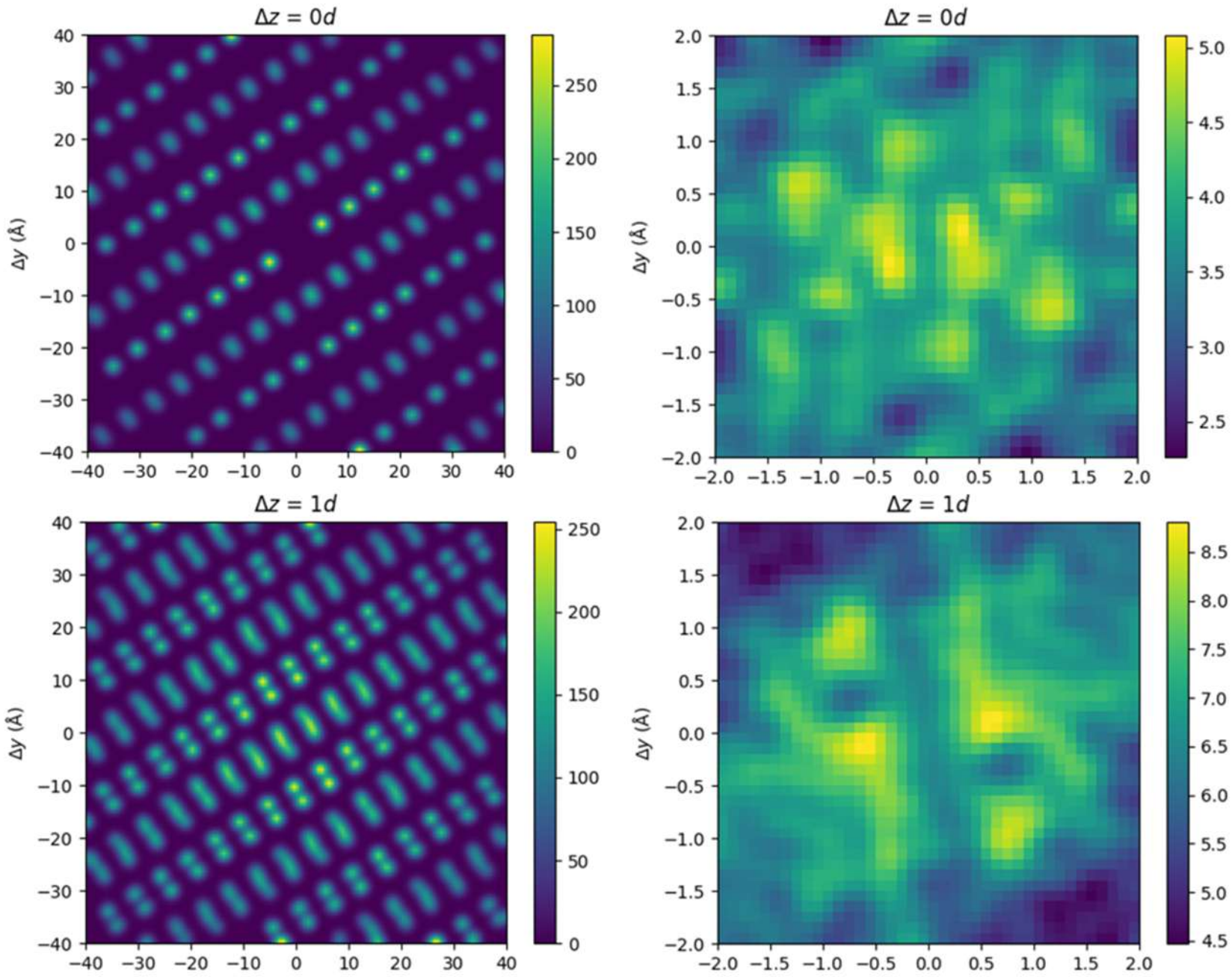

***Figure S10.6**. Comparison between calculated and experimental xy-SDM correlations for the proposed crystalline 8CB orientation. Left: calculated in-plane molecular correlation maps for the proposed crystalline orientation at Δz = 0d and Δz = 1d, where d represents the experimentally measured interplanar spacing of 0.731 nm. Right: corresponding experimental xy-SDM maps extracted from the periodic 8CB region under the same Δz conditions. Due to the limited ion counts, the experimental maps do not allow definitive two-dimensional crystallographic indexing. Nevertheless, weak non-random local correlations are observed, providing supporting evidence for the proposed crystalline molecular periodicity.*

### S10.7 Comparison with reconstruction artefacts

Finally, we compared the periodic molecular feature with a typical stripe-like artefact observed in other regions of the reconstruction. As shown in **Figure S10.7**a, the artefact appears as irregular field-evaporation/reconstruction-related streaks without a well-defined periodic spacing. This morphology is clearly different from the layered molecular feature shown in **Figure S10.7**b, which exhibits a regular one-dimensional spacing and is mainly associated with molecular signals.

This comparison is not used as standalone proof of crystallographic ordering. Instead, it provides an additional control showing that the feature discussed here is not simply periodic instability of the evaporation sequence.

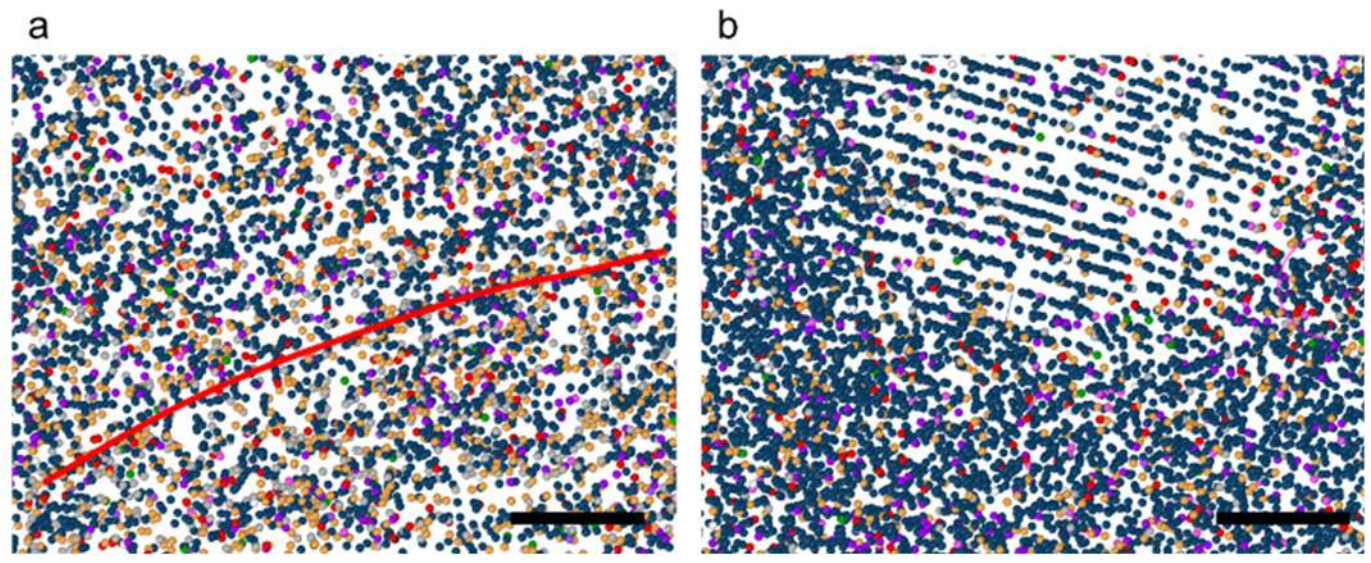

***Figure S10.7***. *Comparison between a periodic instability of the evaporation sequence indicted by the red line and the periodic molecular feature in the 8CB dataset. Scale bar: 5 nm.*

**S10.8 Summary**

Taken together, the periodic molecular feature satisfies several basic criteria commonly used to evaluate plane-like contrast in APT: it is approximately tangent to the local specimen curvature, it is associated with a localized pole-like evaporation feature, it shows a clear z-SDM periodicity, its periodic response is mainly associated with molecular signals rather than background signals, its spacing is inconsistent with smectic layering but consistent with the reported crystalline 8CB structure, and it is morphologically distinct from periodic instability of the evaporation sequence.

We therefore do not interpret this feature as conventional atomic-plane imaging. Instead, we assign it cautiously to molecular crystallographic periodicity in a crystalline 8CB domain. This assignment provides the structural basis for the discussion in the main text regarding the different field-evaporation behaviors of smectic-like and crystalline 8CB regions.

## S11. Correlation between voltage evolution and fragment distributions in heat-treated 8CB and 8OCB datasets

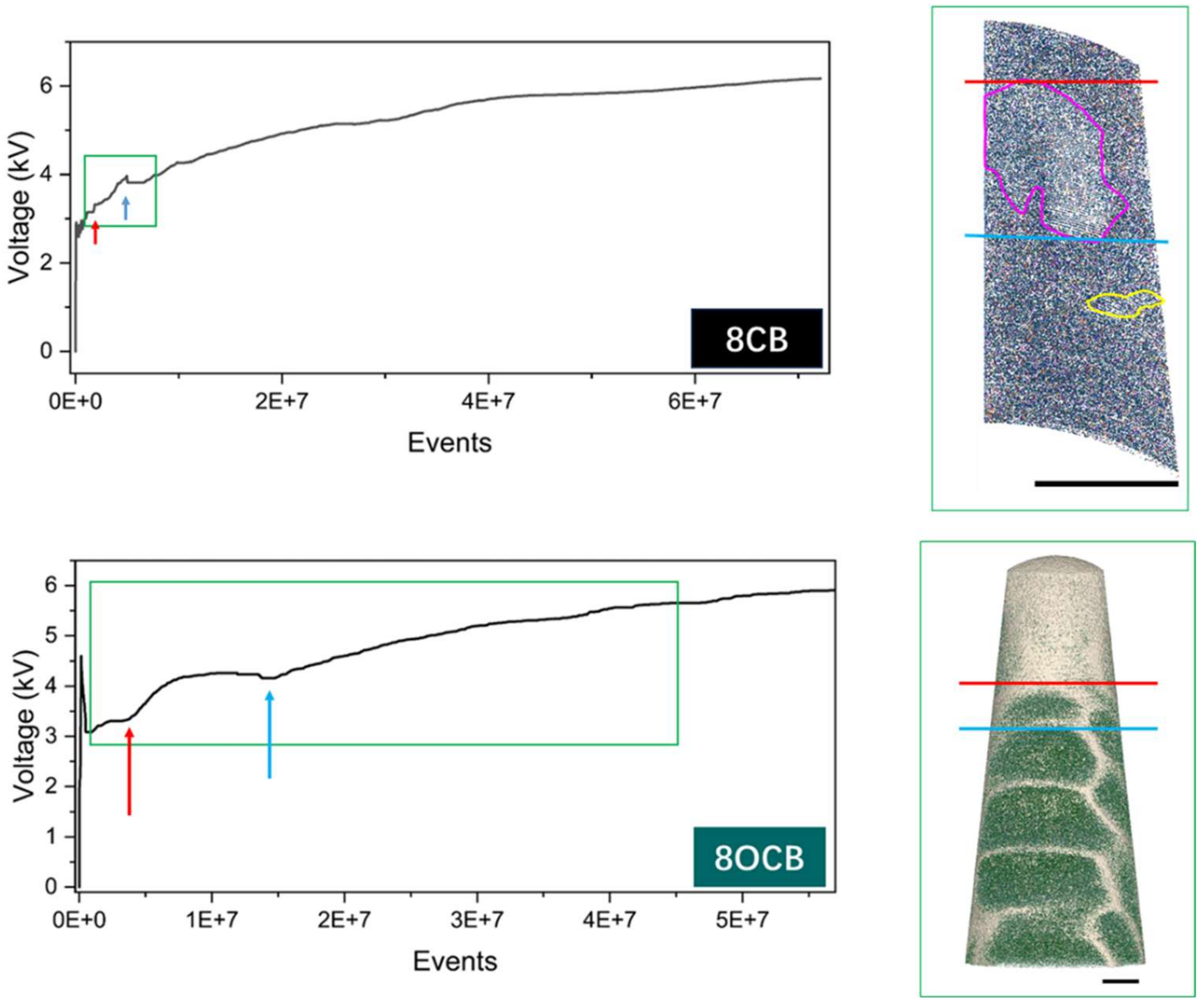

***Figure S11.1***. *Correlation between voltage evolution and event distributions in heat-treated 8CB and 8OCB datasets. The experimental voltage curves of the heat-treated 8CB and 8OCB measurements are compared with the ideal voltage evolution expected under a fixed-shank-angle assumption. In both datasets, the experimental curves show only limited deviation from the ideal trend, supporting the use of a geometry-based reconstruction with a fixed shank angle for these measurements. Colored arrows on the voltage curves indicate selected evaporation stages at different voltage levels, and the correspondingly colored lines in the reconstructed volumes mark their spatial positions along the evaporation sequence. This comparison links the voltage evolution to the reconstructed event distribution and shows that the main reconstructed regions used for analysis are consistent with the expected evaporation process. Black scale bars: 50 nm..*

**S12. Supplementary number-fraction analysis in 8OCB**

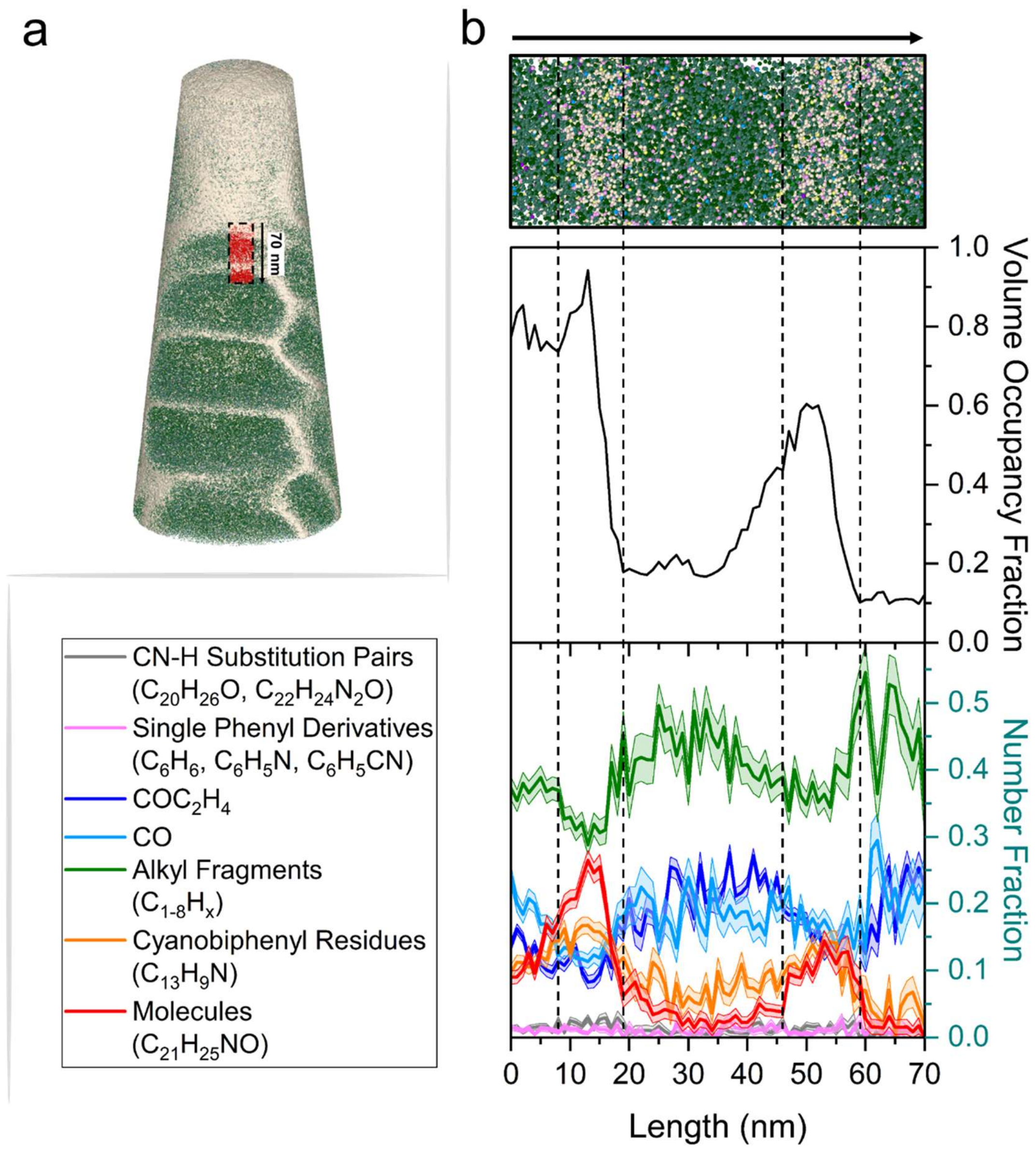

***Figure S12.1**. Visualization of solid-liquid interfaces in pure 8OCB. (a) Reconstructed 8OCB specimen showing crystalline domains embedded in surrounding smectic network. (b) One-dimensional volume occupancy fraction and number fraction profile across the solid-liquid interface highlighted in (a). The analysis was performed over a cylindrical region (10 nm radius, 70 nm length).*

## S13. Other information

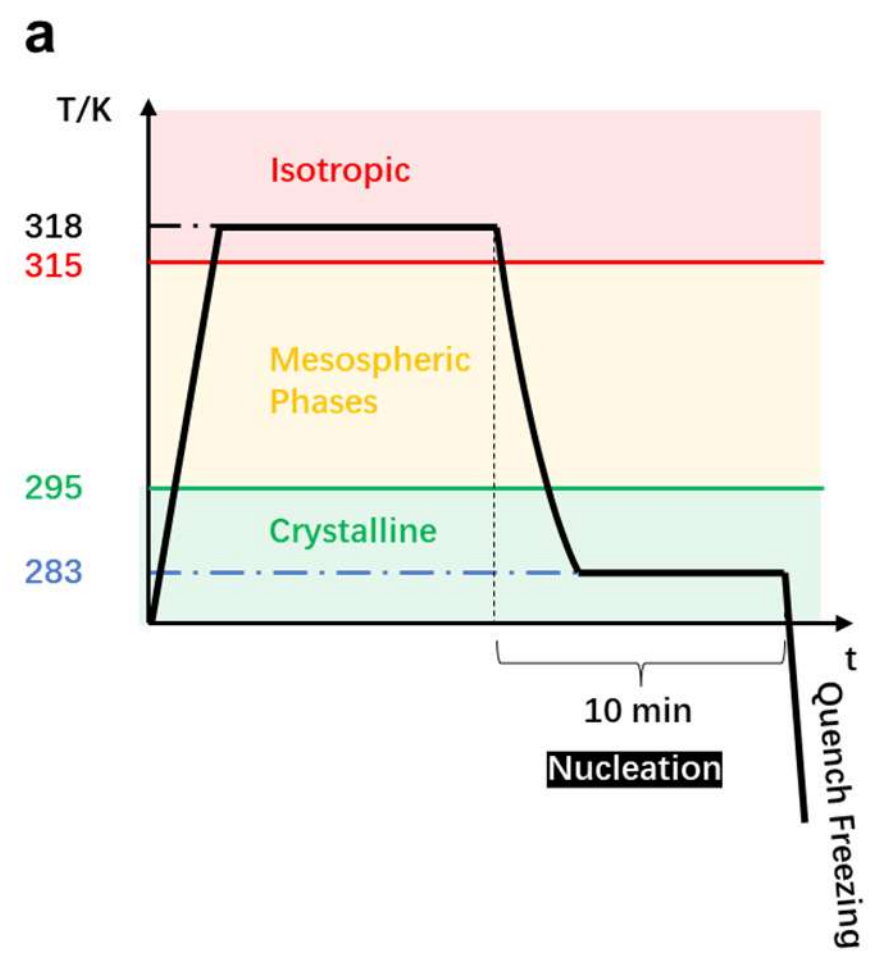

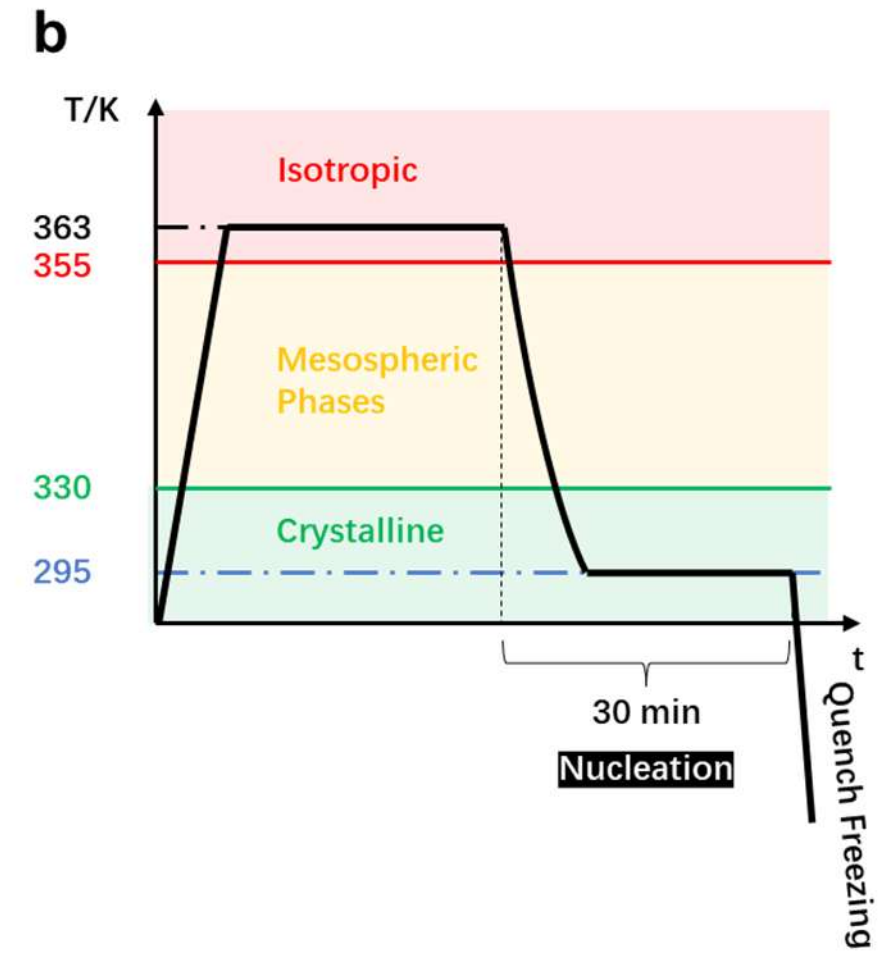

***Figure S13.1***. *Thermal protocols applied to (a) 8CB and (b) 8OCB prior to quench-freezing based on phase diagram of 8CB-8OCB* (Dayal et al., 2006).

In both cases, the LC was deposited into the MicroCup in its isotropic phase. The temperature was then reduced to a value slightly below the respective crystallization temperature and held for a short period to promote controlled nucleation. After this brief annealing step, the samples were rapidly quenched to cryogenic temperatures to preserve the nucleated structures for subsequent FIB milling.

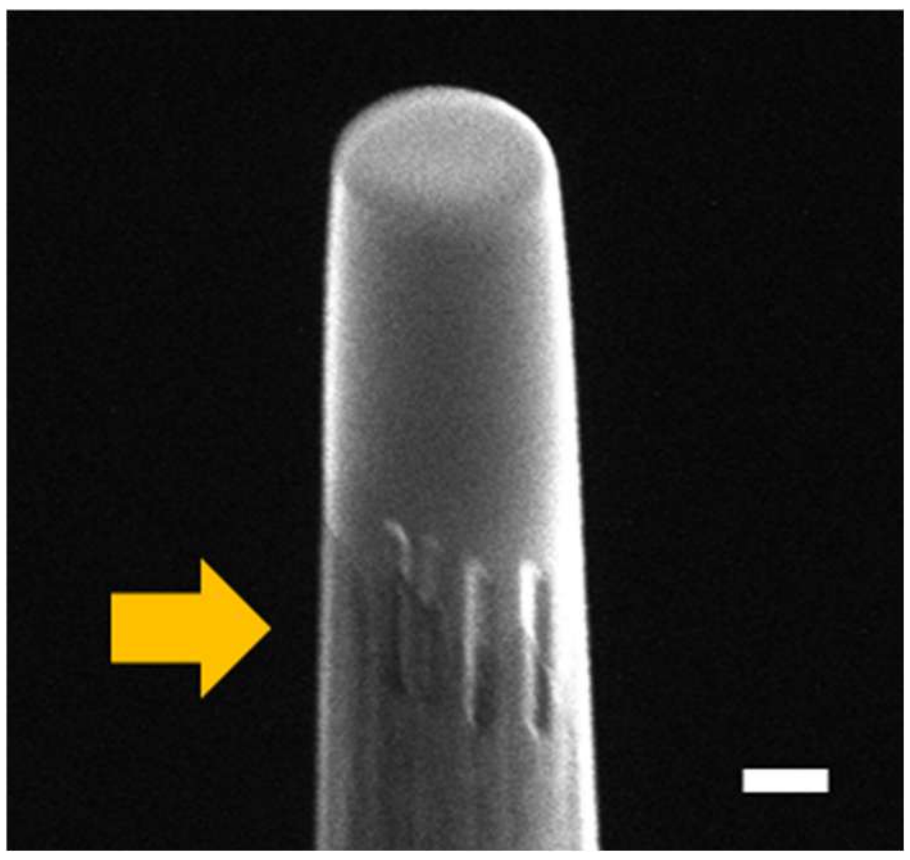

***Figure S13.2***. *Electron-beam image illustrating a redeposition-induced attachment step commonly used in cryo-liftout workflows* (Schreiber et al., 2018; Woods et al., 2023). *Scale bar: 1 µm.*

To validate whether this method could further strengthen specimen adhesion, we applied the same redeposition procedure to a MicroCup-prepared liquid sample. However, no measurable improvement in mechanical stability was observed. Unlike lift-out specimens, where redeposition is required to secure the lamella, organic liquids naturally anchor within the MicroCup cavity during deposition.